\documentclass[preprint,12pt]{elsarticle}

\usepackage{amssymb}
\usepackage{amsmath}
\usepackage{amsthm}
\usepackage[margin=1in]{geometry}
\usepackage{latexsym}
\usepackage{graphicx}
\usepackage{amsfonts}
\usepackage[colorlinks,allcolors=blue]{hyperref}
\usepackage{mathtools}
\usepackage{xcolor}
\usepackage{bbm}

\usepackage{color-edits}

\journal{Computational and Structural Biotechnology Journal}

\usepackage{booktabs}

\begin{document}

\begin{frontmatter}



\title{Assessing the Role of
Volumetric Brain Information in Multiple Sclerosis Progression}

\author[1]{Andy A. Shen\footnote{A. Shen and A. McLoughlin contributed equally to this work.}}
\author[2]{Aidan McLoughlin\footnotemark[1]}
\author[1]{Zoe Vernon}
\author[3]{Jonathan Lin}
\author[4]{Richard A.D. Carano}
\author[1,2]{Peter J. Bickel}
\author[4]{Zhuang Song\footnote{Z. Song and H. Huang are co-last, co-corresponding authors.}}
\author[1,2]{Haiyan Huang\footnote{Correspondence to: 367 Evans Hall, Berkeley, CA 94720; (510) 642-2781 \\ E-mail addresses: \texttt{song.zhuang@gene.com} (Z. Song), \texttt{hhuang@stat.berkeley.edu} (H. Huang)}}

\affiliation[1]{organization={Department of Statistics, UC Berkeley},
            city={Berkeley},
            state={CA},
            country={USA}}
\affiliation[2]{organization={Division of Biostatistics, UC Berkeley},
            city={Berkeley},
            state={CA},
            country={USA}}
\affiliation[3]{organization={Department of Statistical Science, Duke University},
            city={Durham},
            state={NC},
            country={USA}}
\affiliation[4]{organization={Analytics and Medical Imaging, Product Development, Genentech Inc.},
            city={South San Francisco},
            state={CA},
            country={USA}}

\begin{abstract}
Multiple sclerosis is a chronic autoimmune disease that affects the central nervous system. Understanding multiple sclerosis progression and identifying the implicated brain structures is crucial for personalized treatment decisions. Deformation-based morphometry utilizes anatomical magnetic resonance imaging to quantitatively assess volumetric brain changes at the voxel level, providing insight into how each brain region contributes to clinical progression with regards to neurodegeneration. Utilizing such voxel-level data from a relapsing multiple sclerosis clinical trial, we extend a model-agnostic feature importance metric to identify a robust and predictive feature set that corresponds to clinical progression. These features correspond to brain regions that are clinically meaningful in MS disease research, demonstrating their scientific relevance. When used to predict progression using classical survival models and 3D convolutional neural networks, the identified regions led to the best-performing models, demonstrating their prognostic strength. We also find that these features generalize well to other definitions of clinical progression and can compensate for the omission of highly prognostic clinical features, underscoring the predictive power and clinical relevance of deformation-based morphometry as a regional identification tool.
\end{abstract}




\end{frontmatter}



\newcommand\blfootnote[1]{%
  \begingroup
  \renewcommand\thefootnote{}\footnote{#1}%
  \addtocounter{footnote}{-1}%
  \endgroup
}

\blfootnote{\textit{Abbreviations:} MS, multiple sclerosis; MRI, magnetic resonance imaging; DBM, deformation-based morphometry; VBM, voxel-based morphometry; LOCO-MP, leave-one-covariate-out minipatch; RSF, random survival forest; CNN, convolutional neural network; CCDP24, confirmed composite disability progression sustained for at least 24 weeks; T25FW, timed 25-foot walking speed test; 9HPT, 9-hole peg test; EDSS, Expanded Disability Status Scale; S25FW, sustained disability progression in 25-foot walking test sustained for at least 24 weeks; BT25FW, baseline timed 25-foot walking speed test}

\newpage

\newenvironment{revision}{\color[HTML]{000000}}{\ignorespacesafterend}

\newcommand{\smallrevision}[1]{\textcolor[HTML]{000000}{#1}}

\section{Introduction} \label{sec:introduction}

Multiple sclerosis (MS) is a chronic autoimmune disease that affects the central nervous system (CNS) in young adults \cite{COMPSTON20081502}. Magnetic resonance imaging (MRI) of the brain is a key tool for diagnosing and monitoring MS progression. Different types of MRI data possess overlapping yet complementary properties and applications. For instance, brain volume measurements derived from T1-weighted MRI provide insights into brain atrophy, which is associated with disability and disease progression \cite{chard2002brain, bermel2006measurement,lomer2024predictors, fisniku2008gray}. Identifying specific brain regions with atrophy and volume changes that are linked to disease progression enhances patient monitoring and facilitates the evaluation of treatment efficacy \cite{song2022deformation}.
\begin{revision}
    Early efforts to assess volume changes in MS consisted of using voxel-based morphometry (VBM), a method designated for volume assessment of grey matter regions \cite{ashburner2000voxel}. VBM demonstrated limited associations with clinical outcomes in MS studies \cite{lansley2013localized} and a majority of VBM studies were cross-sectional and thus unable to track disease-related changes at the individual level. On the other hand, deformation-based morphometry (DBM)  \cite{ashburner1998identifying,chung2001unified} provides voxel-level measurements of volume assessment across the entire brain, including grey matter, white matter, and the ventricles. We note that enlargement of the fluid-filled ventricles in the brain typically reflect either local or global brain tissue loss, which cannot be captured by VBM due to its focus on grey matter regions. While VBM often suffers from tissue segmentation errors \cite{chung2001unified,manera2019deformation}, DBM only relies on image registration, making it a more robust tool to assess voxel-based volume differences.
\end{revision}

\smallrevision{MRI data has proven instrumental in predicting disease progression and plays a critical role in personalizing patient treatment decisions \cite{rocca2019application,wattjes2015magnims}.} Despite recent progress in using T1-weighted MRI data to predict MS progression \cite{moazami2021machine,vazquez2023systematic,lomer2024predictors,prathapan2024modeling}, several open questions remain regarding how data and methodological choices may affect model reliability. For example, recent work leveraging random survival forests achieved high accuracy in predicting clinical progression over a 10-year period based on changes in T1-weighted MRI-derived features within the first 2 years \cite{pisani2021novel}. However, baseline T1-weighted MRI data alone demonstrated relatively limited predictive power for 2-year progression \cite{pellegrini2020predicting}. Similarly, it was shown that, while voxel clusters derived from independent components analysis (ICA) applied to 2-year volume change data were predictive of 10-year progression, the voxel cluster values at baseline were similar between progressors and non-progressors \cite{bergsland2018gray}. Some studies have also noted unclear predictive value of T1-weighted MRI, which can vary between studies and can depend on outcome definitions \cite{pellegrini2020predicting,tousignant2019prediction}. These issues also persist in deep learning frameworks \cite{zhang2023predicting,coll2024global}. It also remains unclear whether T1-weighted MRI data provides complementary information for progression prediction beyond clinician-measured variables of disease state, some of which are used to define common progression endpoints themselves \cite{sormani2010surrogate,cadavid2017edss}. Ultimately, there is a lack of literature on detecting baseline volume measurements from T1-weighted brain MRI that are predictive of clinical progression within two years. \smallrevision{For broader reviews of data-driven approaches to clinical modeling of MS, we refer readers to \cite{moazami2021machine,vazquez2023systematic,lomer2024predictors,prathapan2024modeling,yousef2024predicting}.}

In this study, we utilize \smallrevision{baseline} deformation-based morphometry derived from T1-weighted MRI of a Phase III MS clinical trial \cite{hauser2017ocrelizumab,song2022deformation} to investigate the impact of various methodological choices, particularly those used to identify predictive DBM features, on the reliability of conclusions drawn from this dataset. We first employ a state-of-the-art, model-agnostic feature importance algorithm, adapted using leave-one-covariate-out methods \cite{lei2014distribution} and minipatch ensembles (LOCO-MP) \cite{gan2023modelagnosticconfidenceintervalsfeature}, to identify brain regions from low-signal T1-weighted MRI baseline data that are predictive of MS progression. \smallrevision{Broadly speaking, LOCO-MP computes feature importance by aggregating predictions from models trained on multiple small, randomly selected subsets of the data (minipatches).} 
\begin{revision}
    We observe that LOCO-MP combined with a survival random forest model identifies features in a more consistent and stable manner than survival random forests alone. This is because LOCO-MP leverages learning from small feature subsets to isolate the effects of highly correlated regions while other methods often yield inconsistent results when the data are highly correlated and high-dimensional \cite{agarwal2023mdiflexiblerandomforestbased}.
\end{revision}


Next, we apply traditional survival models to examine how different outcome/endpoint definitions, in conjunction with feature selection by LOCO-MP, influence prediction power. Although our progression endpoint (CCDP24, discussed in Section \ref{sec:data}) is highly censored within the two-year study period, it represents a well-defined and clinically meaningful composite marker of MS progression. To validate the DBM features identified using LOCO-MP, we assess their generalizability across alternative progression endpoints (S25FW, discussed in Section \ref{sec:data}) to compare a single, objective measure of physical disability with the composite measure, the latter involving some degree of clinician discretion \cite{sormani2010surrogate}. \smallrevision{We also verify that our selected features generalize well across a separate patient cohort (see \ref{sec:generalizability}).} 

By using DBM at baseline to identify patient volumetric abnormalities and by applying LOCO-MP to enforce sparsity in the DBM feature set, we achieve improved prediction performance compared to models that use whole-brain T1-weighted MRI data as input. \smallrevision{Additionally, we compare the model performance of DBM features against models using conventional, non-DBM clinical features (such as age and sex), also measured at baseline. We introduce these conventional features in Section \ref{sec:data}.} Interestingly, models using selected MRI features alone perform on par with those using only the \smallrevision{baseline} conventional measurements. Combined models incorporating both selected MRI features and conventional measures do not necessarily outperform individual models (those using either selected MRI features or conventional measures alone). Notably, removing \smallrevision{the conventional features} that define certain endpoints significantly reduces the predictive power of the conventional feature models, underscoring the strong predictive value of the selected MRI features. 

We further analyze whether 3D convolutional neural networks (CNNs) using only the LOCO-MP regions perform better than full brain architectures. We develop a 3D CNN architecture, \textit{Region CNN}, that accepts a set of atlas region tensors as input. While full brain MRI deep transformer models \cite{coatnet, yu2023unest} are unstable and show low prediction accuracy, \textit{Region CNN} using LOCO-MP identified features performs substantially better. This highlights the importance in filtering out unnecessary regions via LOCO-MP for extracting prognostic signal from CNNs for this cohort.


\section{Materials and methods} 
\label{sec:methodology}
\subsection{Data and image preprocessing} \label{sec:data}
The data used in this study are from a comparator arm of a phase 3 clinical trial of relapsing multiple sclerosis (OPERA I: NCT01247324), in which the patients were treated with interferon (IFN) $\beta$-1a (44 $\mu$g) three times per week throughout the 96-week treatment period. Our sample consisted of 350 patients. Details on patient selection, MRI acquisition, and clinical assessments are provided in the original report \cite{hauser2017ocrelizumab}. Briefly, patients were recruited with an age range of 18 to 55 years; McDonald criteria diagnosis of multiple sclerosis; Expanded Disability Status Scale (EDSS) score between 0 to 5.5; more than 2 documented clinical relapses within the previous 2 years or one clinical relapse within the year before screening; brain MRI evidence of multiple-sclerosis-related abnormalities. Conventional T1-weighted 3D spoiled gradient-recalled echo brain MRI was acquired at baseline, Weeks 24, 48 and 96 (repetition time = 28–30 ms, echo time = 5–11 ms, flip angle = 27–30 deg, 60 oblique axial slices of 1 mm in-plane resolution and 3 mm slice thickness). The primary clinical outcome was defined as a composite measure of disability progression which included three clinical assessments: EDSS, Timed 25-Foot Walk (T25FW, a measure of short distance walking speed), and 9-Hole Peg Test (9HPT, a measure of upper limb function) \cite{elliott2019chronic}. This composite measure was developed to capture a broader aspect of disability in patients with multiple sclerosis. The 24-week composite confirmed disability progression (CCDP24) was defined as progression on any one of the three components (EDSS, T25FW, or 9HPT): an increase of EDSS score from the baseline at least 1.0 point (or 0.5 points if the baseline EDSS score was larger than 5.5) or a 20\% minimum threshold change for T25FW and 9HPT. We also consider a 24-week progression outcome based solely on the 25-foot walking score (S25FW) in Section \ref{sec:s25fw-results}. \smallrevision{Table \ref{tab:data-summary} summarizes the conventional features we use for prediction, grouped by those who experienced CCDP24 progression and those who were censored.}

Since MS progression is influenced by many factors, our models consider a variety of demographic and clinical measurements as well as standard imaging metrics. Broadly speaking, we refer to these features as ``conventional features'' \smallrevision{throughout the manuscript} to distinguish them from DBM-based features.  
The conventional features serve as a comparison against the DBM features to assess which feature set performs more optimally.  We record nine conventional features which include the patient's age, birth sex, years of onset, weight (in kg), total brain volume, total T2 lesion volume\smallrevision{\footnote{The T2-weighted MRI sequence was described by Elliot et al. (2019) \cite{elliott2019chronic}: axial 3mm T2- weighted slices were acquired with 2D fast spin-echo, repetition time = 4000–6190 ms, echo time = 74–91 ms and echo train length = 7–11. The T2 lesion volume was measured from T2-weighted MRI images using a proprietary, semi-automatic method. The initial automatic segmentation was performed using a Bayesian classifier \cite{francis2004automatic} which was then verified and corrected by qualified readers.}}, baseline 25-foot walking test score, baseline 9-hole peg test score, and baseline EDSS, which are generally thought to be related to MS disease progression \cite{bar2023blood}.

\begin{table}[ht]
\centering
\caption{Baseline characteristics stratified by CCDP24 outcome (censored/progressed)}
\small
\resizebox{\textwidth}{!}{%
\begin{tabular}{lccc}
\toprule
\textbf{Variable} & \textbf{Censored (N = 269)} & \textbf{Progressed (N = 81)} & \textbf{p-value} \\
 & Mean (SD) or n / N (\%) & Mean (SD) or n / N (\%) & \\
\midrule
Baseline age (years)                 & 37 (9)         & 37 (10)        & $> 0.9$   \\
Years of onset                       & 6.3 (6.2)      & 7.3 (6.0)      & 0.088  \\
Baseline weight (kg)                & 76 (18)        & 74 (16)        & 0.4    \\
Baseline 9-hole peg test score time (seconds)      & 23.1 (6.6)     & 24.2 (5.2)     & 0.033  \\
Baseline 25-foot walking test score time (seconds) & 7.0 (5.0)      & 5.9 (3.7)      & 0.049  \\
Baseline EDSS                        & 2.73 (1.28)    & 2.57 (1.25)    & 0.4    \\
T2 lesion volume                     & 8 (9)          & 13 (14)        & 0.019  \\
Brain volume                         & 1,501 (86)     & 1,480 (85)     & 0.066  \\
Proportion female                    & 185 / 269 (69\%) & 50 / 81 (62\%) & 0.2    \\
\bottomrule
\end{tabular}
}
\vspace{1ex}
\raggedright
\textsuperscript{Wilcoxon rank sum test for continuous variables; Pearson’s Chi-squared test for categorical variable.}
\label{tab:data-summary}
\end{table}


Statistical features of regional brain volume were extracted from T1-weighted brain MRI, using a deformation-based morphometry (DBM) pipeline based on the diffeomorphic image registration of advanced normalization tools \cite{tustison2019longitudinal}. 
\begin{figure}[ht]
    \centering
    \includegraphics[width=0.9\linewidth]{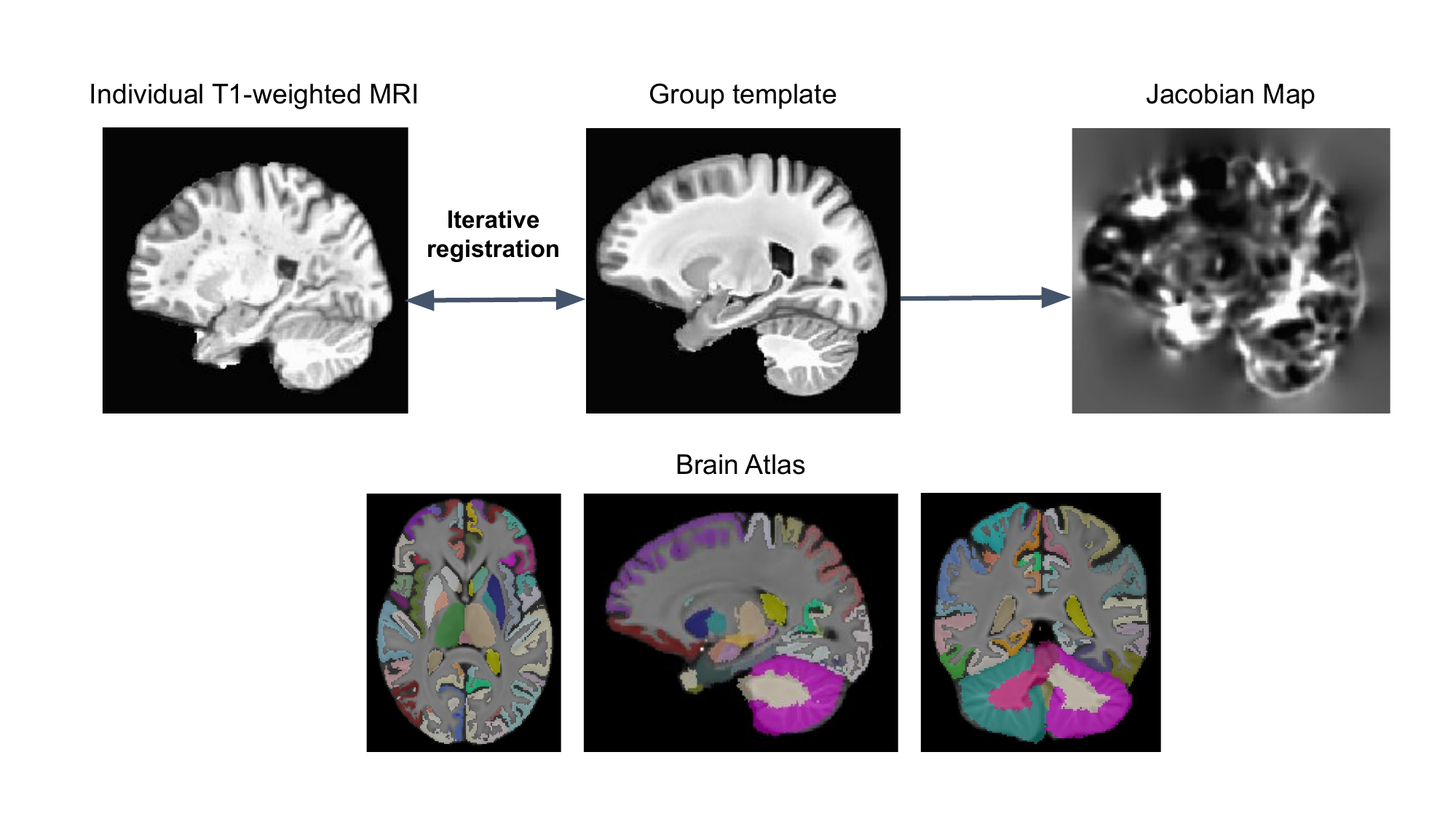}
    \caption{Illustration of the deformation-based morphometry (DBM) analysis. For each T1-weighted brain MR image, the registration with the group template generates a Jacobian map over the whole brain, which is a voxel-level measure of volume relative to the group template. Regional labels on the brain atlas were mapped to the group template to extract statistical features of the Jacobian maps at the patient level.}
    \label{fig:dbm-schematic}
\end{figure}
The Mindboggle atlas was used to identify individual brain regions \cite{klein2017mindboggling}. In addition, prior probability images of six major brain regions based on the Mindboggle atlas, including CSF, cortical grey matter, deep grey matter, cortical white matter, the brainstem, and the cerebellum, were binarized at a probability threshold of 0.5. A population-specific group template of the T1-weighted brain MRI was constructed by advanced normalization tools with T1-weighted brain MRI images of 171 healthy adults (aged 20–59 years). Before being fed into the pipeline, each individual brain image  was preprocessed with the following steps: 1) resampling to an isotropic resolution of 1 mm; 2) N4 bias correction \cite{tustison2010n4itk}; and 3) denoising with a non-local algorithm \cite{manjon2010adaptive}. Regional labels were mapped from the group template to individual images so that volume change of any predefined region could be measured. 

\smallrevision{As mentioned in the introduction, all of the conventional and DBM features we use for prediction are measured at baseline.} 
The tabular DBM feature set consists of the median and standard deviation voxel values for each of the 51 atlas regions, resulting in 102 total features. Prior to performing feature selection and modeling, we filter out features with a variance below 0.01 across all patients, reducing the number of DBM features from 102 to 56. No other conventional features were excluded in this filtering step. 

\subsection{Feature importance using minipatch ensembles} \label{sec:loco-mp}
\begin{figure}[ht]
    \centering
    \includegraphics[width=0.95\linewidth]{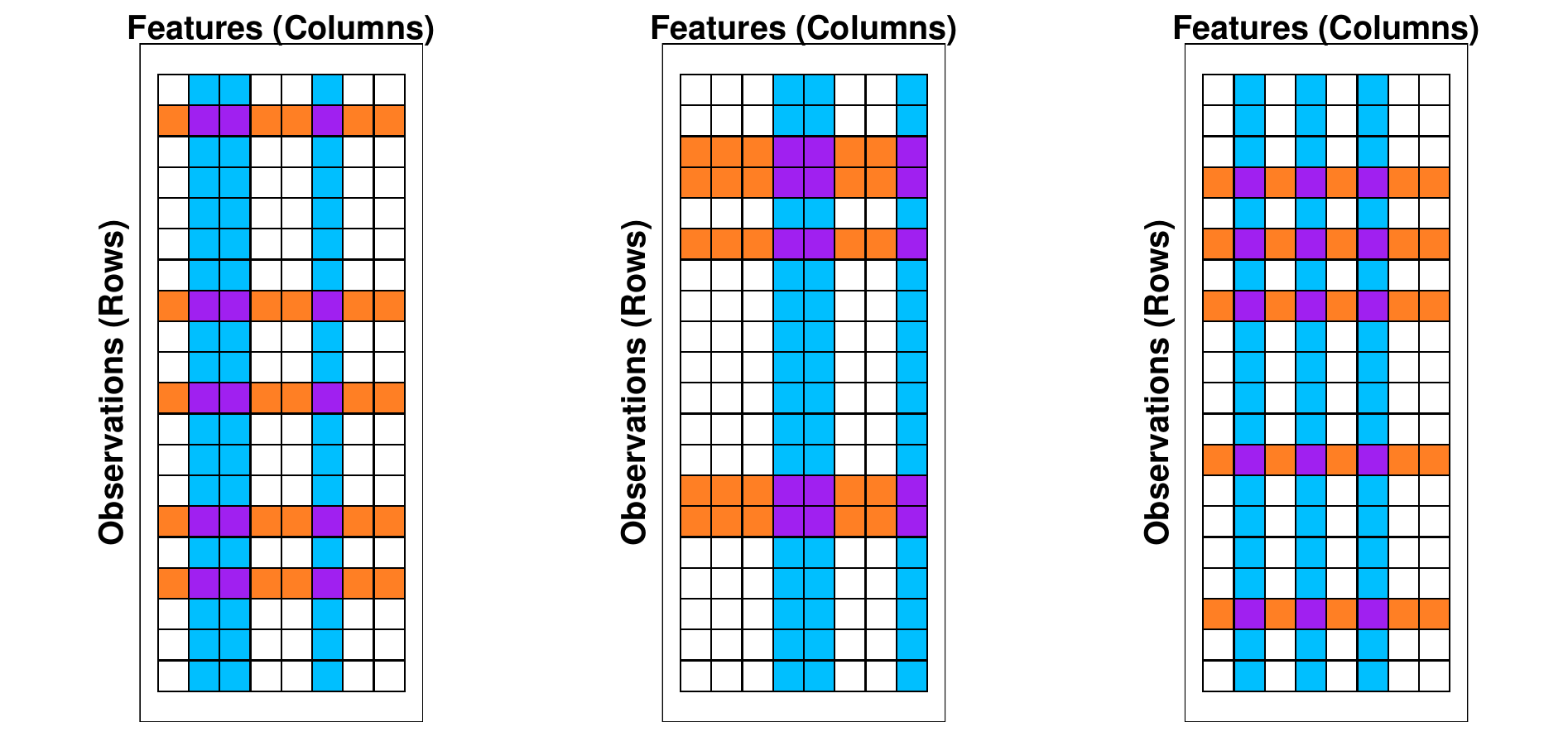}
    \caption{General overview of LOCO-MP. Each plot demonstrates how a subset of observations (orange) and features (blue) are subsampled to form a ``minipatch'' (purple cells). A model is trained on the purple cells and predictions are generated using the unselected observations (blue cells). This process is repeated over numerous minipatches to ensure each feature and observation is selected enough times to perform inference. Feature importance for a target feature is assessed by evaluating performance on minipatches with and without that feature.}
    \label{fig:loco-schematic}
\end{figure}

\begin{revision}
    Identifying features that genuinely contribute to prediction is crucial for model interpretability and guiding practical decision-making. Feature selection methods are used to filter out variables that are inconsequential to prediction, retaining those with the most predictive value. Since our dataset consists of high-dimensional and highly correlated features, traditional feature selection methods can potentially misattribute the importance of a feature when other correlated features are present. To offer a more reliable assessment of feature importance, we extend a feature importance method called \textit{leave-one-covariate-out minipatch (LOCO-MP)} prediction \cite{lei2014distribution,yao2021featureselectionhugedata,gan2022fast,gan2023modelagnosticconfidenceintervalsfeature} to survival analysis. LOCO-MP consists of taking a subsample of the data, called a ``minipatch.'' The purple cells in Figure \ref{fig:loco-schematic} shows a simple example of what three minipatches may look like. A model is then trained on the minipatch and this process is repeated over many minipatches to ensure an even distribution of observations and features are sampled. After training on all minipatches, LOCO-MP determines feature importance for a certain target feature $j$ by comparing the average prediction error across minipatches with and without feature $j$. 

    LOCO-MP is advantageous over other feature selection techniques due to its ability to provide asymptotic inferential guarantees  \cite{gan2023modelagnosticconfidenceintervalsfeature}. Moreover, LOCO-MP can also account for dependencies across features, a common issue in high-dimensional settings like ours --  Gan et al. (2023) \cite{gan2023modelagnosticconfidenceintervalsfeature} discuss that, by generating randomly subsampled features across minipatches, LOCO-MP ensures that the predictive value of each feature is not being diminished by other strongly correlated features, since groups of strongly correlated features will not always appear in the same minipatch. The full LOCO-MP procedure and its theoretical guarantees are discussed in Gan et al. (2023) \cite{gan2023modelagnosticconfidenceintervalsfeature}.

    The procedure is summarized graphically in Figure \ref{fig:loco-schematic} and outlined below for a target feature $j$: 
    \begin{enumerate}
    \item \textbf{Subsample the data:} Subsample a set of observations (rows) and features (columns), denoted by purple cells. This forms a ``minipatch.''
    \item \textbf{Train a model:} Fit a model to the minipatch and evaluate predictions using the remaining observations (blue cells).
    \item Repeat steps 1 and 2 over $K$ minipatches, where $K$ is user-determined.
    \item \textbf{Compute feature importance:} Compute $\Bar{\Delta}_{j}$, the feature importance score of feature $j$ (discussed in the following paragraphs). 
\end{enumerate}
\end{revision}

\begin{revision}
Our outcome of interest, patient progression, is recorded as a time-to-event pair $(T_i, C_i)$, where $T_i$ is the event time and $C_i \in \left\lbrace 0,1 \right\rbrace$ is an indicator for whether the patient's disease progression is observed during the study period ($C_i = 1$) or is right-censored ($C_i = 0$).
Because our outcomes of interest are recorded in this time-to-event format, we extend LOCO-MP to survival analysis. This is done by predicting discrete instantaneous hazards from the trained models and computing an associated individual-level discrete hazard loss~\cite{zadeh2020bias, vale2021long}. We refer the reader to \ref{sec:loco-details} for a full technical description of LOCO-MP and the hazard loss function used to characterize prediction error in our survival models.
\end{revision}

\newcommand{\XIkFk}{X_{I_k, F_k}}
\newcommand{\XminusIkFk}{X_{{-I_k, F_k}}}
\newcommand{\mukhat}{\hat{\mu}_k}


\subsection{MS progression with classical survival models} \label{sec:classical-survmod}
To assess the predictive power of the DBM and conventional features, we apply random survival forests (RSF) \cite{10.1214/08-AOAS169} to train the tabular dataset on the progression endpoint(s). RSF is an extension of Breiman's random forest algorithm to the time-to-event setting. In \ref{sec:cox-models}, we include additional results using a penalized Cox proportional hazards model with a $\ell_2$ penalty (analogous to ridge regression) \smallrevision{and a $\ell_1$ penalty (analogous to lasso regression)}. 
All analyses in this section were conducted using R (version 4.2.1). RSF models were implemented using the package \texttt{randomforestSRC} \cite{ishwaran2019fast} and the Cox models were implemented using the package \texttt{glmnet} \cite{friedman2021package}.

In order to evaluate the quality of predictive information yielded via LOCO-MP, we trained the survival models on several distinct feature groupings of conventional and DBM features, outlined below:
\begin{enumerate}
    \item \textbf{Conventional-only:} this model consists solely of the conventional features introduced in Section \ref{sec:data}. 
    \item \textbf{All DBM:} this model consists of \textit{all} DBM features (median and standard deviation of each region), after removing features with low variance ($<0.01$). No conventional features are included.
    \item \textbf{Conventional + All DBM:} this model uses all features from the ``Conventional-only'' and ``All DBM" feature sets.
    \item \textbf{Top DBM:} this model consists only of the top selected DBM features from the LOCO-MP algorithm, ranked by their feature occlusion scores $\Bar{\Delta}_j$\footnote{For illustration, we use the top six LOCO-MP features and discuss the motivation for this choice in the results section. We later demonstrate that the specific number of features used for modeling is less important than identifying a broad set of features for prediction.}. No conventional features are included.
    \item \textbf{Conventional + Top DBM:} this model uses all conventional features and includes the top selected DBM features from the LOCO-MP algorithm, ranked by their feature occlusion scores $\Bar{\Delta}_j$.
\end{enumerate}
The first three feature groupings assess the individual and combined performance of the conventional and DBM features. The last two feature groupings are designed to evaluate the predictive potential of the DBM features selected by LOCO-MP. 

The models were trained on each grouping using six repeats of 5-fold cross-validation. Model performance was measured by computing the Harrell's Concordance index (C-index). The C-index is analogous to the area under the ROC curve (AUROC) for binary classification tasks by comparing the observed time-to-event with the predicted risk and computing the proportion of patients where the two values are consistent (e.g. predicting higher risk for a patient with observed progression prior to another patient). A C-index of 1 denotes perfect model discrimination between patients, and a C-index of 0.5 indicates a fully random model. The primary outcome of our analysis is CCDP24, discussed in Section \ref{sec:data}. In Section \ref{sec:survmod-results}, we consider an alternative definition of progression based on the 25-foot walking test (S25FW).

\subsection{MS progression with 3D convolutional neural networks} \label{sec:3d-cnn}
\begin{figure}[ht]
    \centering    \includegraphics[width=0.85\linewidth]{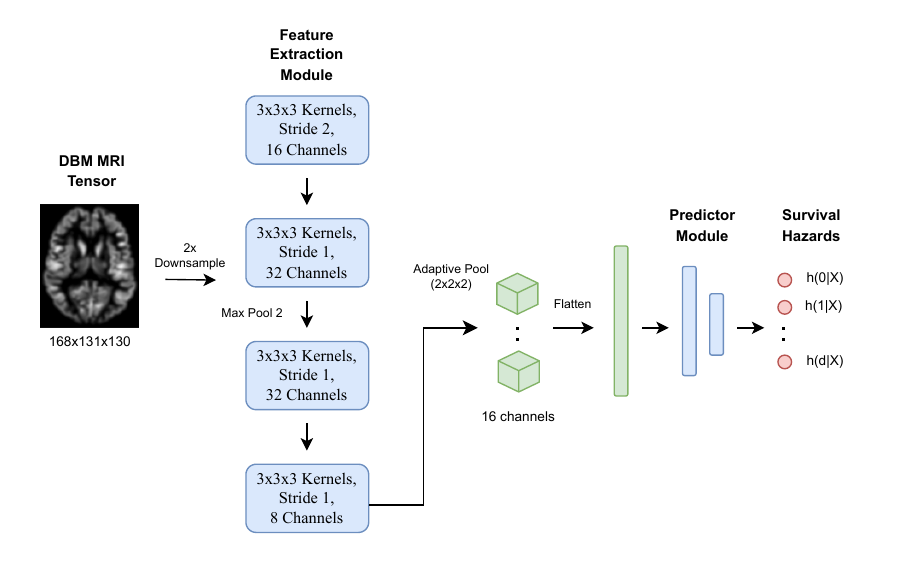}
    \caption{Architecture diagram of the shallow full brain 3D CNN model.  3D convolutional layers extract local volumetric features from the MRI DBM voxel tensor. Summary features from 8 sections of the tensor are collected and passed to a feed-forward predictor module to output the discrete conditional hazards.}
    \label{fig:fb_cnn_diag}
\end{figure}

\begin{figure}[ht]
    \centering    \includegraphics[width=0.85\linewidth]{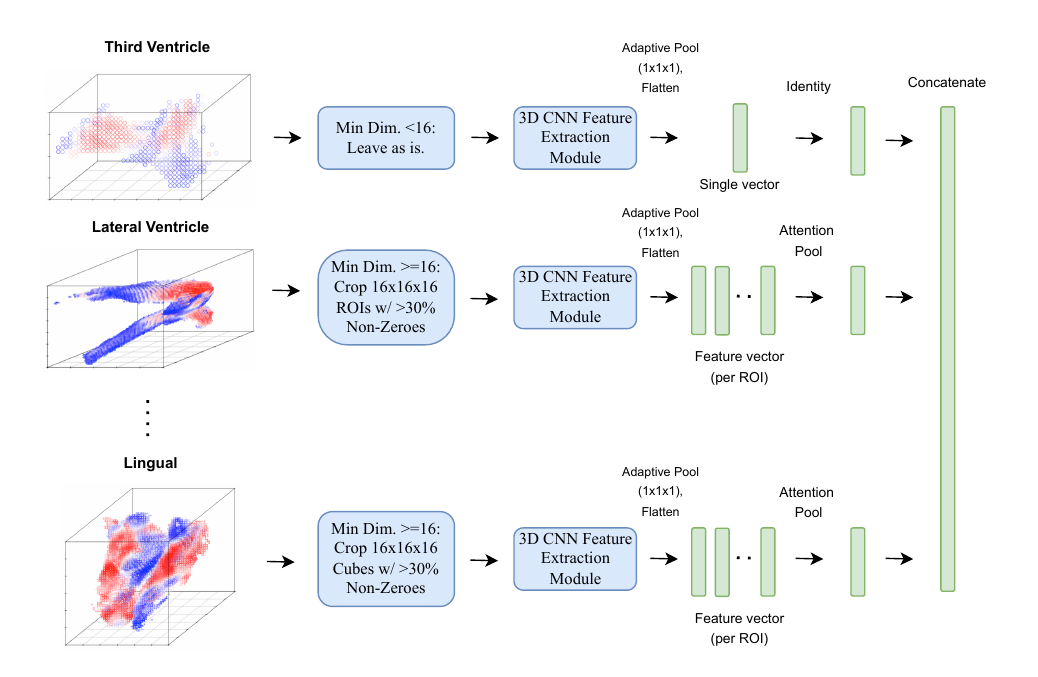}
    \caption{Architecture diagram of the region-based 3D CNN feature extraction. For each atlas region, the DBM voxels are cropped into $16\times16\times16$ regions of interest (ROIs) comprising at least 30\% nonzero values, if the region is larger than this crop size. The ROIs are passed through a 3D CNN feature extractor, following the specs of the feature extraction module in Figure \ref{fig:fb_cnn_diag}. The resulting tensors are spatially pooled and flattened to a feature vector for each ROI. The ROI feature vectors are pooled to a region-level feature vector using attention. Finally, the region-level feature vectors are concatenated to form the final DBM feature vector passed into the survival hazard predictor module from Figure \ref{fig:fb_cnn_diag}. A separate feature extractor is learned jointly for each selected atlas region.}
    \label{fig:reg_cnn_diag}
\end{figure}
In contrast to the summary statistic featurization of the DBM data, 3D CNNs allow for flexible learning of localized features from the full brain or atlas region voxel-level tensors. Several applied works in MS have considered shallow 3D CNNs \cite{coll2024global, tousignant2019prediction, storelli2022deep}. These relatively ``parsimonious'' architectures can be better suited to tasks with small sample size and without pretrained neural network weights to initialize the model with.

Our 3D CNN survival analysis pipeline, as applied to the full brain DBM tensor, is shown in Figure \ref{fig:fb_cnn_diag}. After the original full brain tensor is downsampled, it is passed through four 3D convolutional layers with kernel size 3, followed by a spatial adaptive max pool. The features resulting from the pool are flattened to a vector and passed through a two layer fully connected predictor module, which outputs discrete survival hazards. The network is trained using the same log-likelihood loss function as described in Equation \ref{eqn:log_lik_loss}.  All neural networks are trained using \texttt{Pytorch 2.3.0} and \texttt{Python 3.10.14}.

We also train two additional CNN models from the literature, a deep convolutional attention model (CoAtNet), and a hierarchical transformer model designed for medical segmentation tasks (UNesT) \cite{coatnet, yu2023unest}. A 3D implementation of CoAtNet is used \cite{timm3d} and it is pretrained using ImageNet \cite{imagenet}. For UNesT, the model is pretrained using T1-w MRIs. For both models, the final layer of their feature extractor modules are fed to an analogous predictor module as the in-house shallow architecture in Figure \ref{fig:fb_cnn_diag}.


Finally, we propose a novel convolutional module to process a user selected subset of the atlas regions. This region-level convolutional model, which we refer to as \textit{Region CNN}, permits the user to use domain knowledge or previous feature selection work to select a sparse set of regions for the CNN to completely focus on, rather than processing the full brain.
The architecture for \textit{Region CNN} is depicted in Figure \ref{fig:reg_cnn_diag}. The user first specifies which regions are included in the model and separate CNN weights are then learned for each region. If the region is larger than $16\times16\times16$, it is cropped into regions of interest (ROIs) of this size, only keeping ROIs with sufficient ($>30\%)$ nonzero voxels. Otherwise, the region tensor is left as a single ROI. The region ROIs are passed through an analogous set of four 3D CNN layers as described in Figure \ref{fig:fb_cnn_diag}. The only difference is that the stride of all CNN layers is 1. The resulting ROI feature vectors are collapsed to a single region-level feature vector using attention pooling \cite{attention}. Finally, the region-level feature vectors are concatenated to form the DBM feature vector for the patient.  For model training, this vector is processed through a predictor module in the same fashion as Figure \ref{fig:fb_cnn_diag}.

\section{Results} \label{sec:results}
\smallrevision{We begin by discussing our feature selection results from LOCO-MP and compare this against a traditional random survival forest model (Section \ref{sec:feat-imp}). Sections \ref{sec:survmod-results} and \ref{sec:cnn-results} discuss the prediction modeling from traditional survival models and convolutional neural networks, respectively.}

\subsection{LOCO-MP feature selection} 
\label{sec:feat-imp}
LOCO-MP identifies the most important DBM features for prediction in a stable manner. Each minipatch has dimension $n = N / 5$ and $m = \sqrt{M}$, where $N = 350$ and $M = 56$. LOCO-MP was first applied to the entire dataset with $K=10000$ minipatches. RSF and CCDP24 were used as the prediction model and progression outcome, respectively. The first four columns of Table \ref{tab:loco_top_regions} show the regions, summary statistics (i.e., the features), feature importance scores $\bar{\Delta}_j$, and rankings of the top six features identified from applying LOCO-MP to the full dataset. Figure \ref{fig:rfgain-all-feats} in the Appendix shows this for all 56 features. In the following paragraphs, we demonstrate the robustness of these feature importance results through various stability checks.

\begin{table}[ht]
\footnotesize
\centering
\begin{tabular}{rllcccc}
\toprule
& \textbf{Region} & \multicolumn{1}{c}{\textbf{Feature Summary}} & \multicolumn{2}{c}{\textbf{Full Data}} & \multicolumn{2}{c}{\textbf{Ten Subsamples}} \\ 
\cmidrule(lr){4-5} \cmidrule(lr){6-7}
& & \multicolumn{1}{c}{\textbf{Statistic}}& \textbf{Rank} & $\Bar{\Delta}_j$ & \textbf{Median Rank} & \textbf{Median $\Bar{\Delta}_j$} \\ 
\midrule
& 3rd Ventricle & \multicolumn{1}{c}{Std. Dev} & 1 & 0.00104 & 2 & 0.00090 \\ 
& Lateral Ventricle & \multicolumn{1}{c}{Median} & 2 & 0.00101 & 1 & 0.00095 \\ 
& Precuneus & \multicolumn{1}{c}{Median} & 3 & 0.00088 & 3 & 0.00078 \\ 
& Cerebellum White Matter & \multicolumn{1}{c}{Median} & 4 & 0.00056 & 4 & 0.00064 \\ 
& Lingual & \multicolumn{1}{c}{Median} & 5 & 0.00051 & 5 & 0.00061 \\ 
& Parahippocampal & \multicolumn{1}{c}{Median} & 6 & 0.00051 & 6 & 0.00042 \\ 
\bottomrule
\end{tabular}
\caption{Top six LOCO-MP selected features based on full data and ten subsamples. The feature occlusion scores ($\Bar{\Delta}_j$) and full data/median subsample ranks are also shown.}
\label{tab:loco_top_regions}
\end{table}


\paragraph{Stability of LOCO-MP} In addition to the theoretical guarantees enjoyed by the LOCO-MP framework, we empirically validate its stability by testing its robustness against various data perturbations \cite{yu2020veridical,yu2024veridical}. We first assess the robustness of the feature importance list by generating ten 80\% subsamples of the data and applying LOCO-MP to each subsample, collecting the corresponding feature importance scores ($\bar{\Delta}_j$) and rankings. This is shown in the latter two columns of Table \ref{tab:loco_top_regions}. Despite a low signal-to-noise ratio in the DBM data, the top six features are identical between the full data and subsampled data models. The first two features are reversed in rank, but the difference in magnitude is extremely small (less than $0.00003$).


We measure the consistency of the feature rankings by computing the \textit{Jaccard similarity index} $J$ between the ranks derived from the full dataset and those from the subsamples. This measures the proportion of overlap between two sets, ranging from 0 (no overlap) to 1 (complete overlap). By calculating $J$ for varying numbers of top-ranked features, we aimed to quantify how reliably the top features appeared in the subsampled rankings. Table \ref{tab:jaccard_summary} presents the results for the top 11 features from the full data model. Across the ten subsamples, the average and median values of $J$ decrease after six features before starting to plateau after ten. Therefore, we choose six as a heuristic number of DBM features to include in our model, though we demonstrate in Section \ref{sec:survmod-results} that our results remain relatively robust to varying numbers of top selected features.

\begin{table}[h]
\footnotesize
\centering
\begin{tabular}{ccccccccccc}
\toprule
\multicolumn{1}{c}{\textbf{Number of top features}} & 2 & 3 & 4 & 5 & 6 & 7 & 8 & 9 & 10 & 11 \\ 
\midrule
Average \( J \) & 0.93 & 0.85 & 0.84 & 1.00 & 0.94 & 0.80 & 0.89 & 0.77 & 0.72 & 0.68 \\ 
Median \( J \) & 1.00 & 1.00 & 1.00 & 1.00 & 1.00 & 0.75 & 0.89 & 0.80 & 0.67 & 0.69 \\ 
\bottomrule
\end{tabular}
\caption{Summary of Jaccard index \( J \) for increasing numbers of top features.}
\label{tab:jaccard_summary}
\end{table}

\begin{figure}[h]
    \centering
    \includegraphics[width=0.9\linewidth]{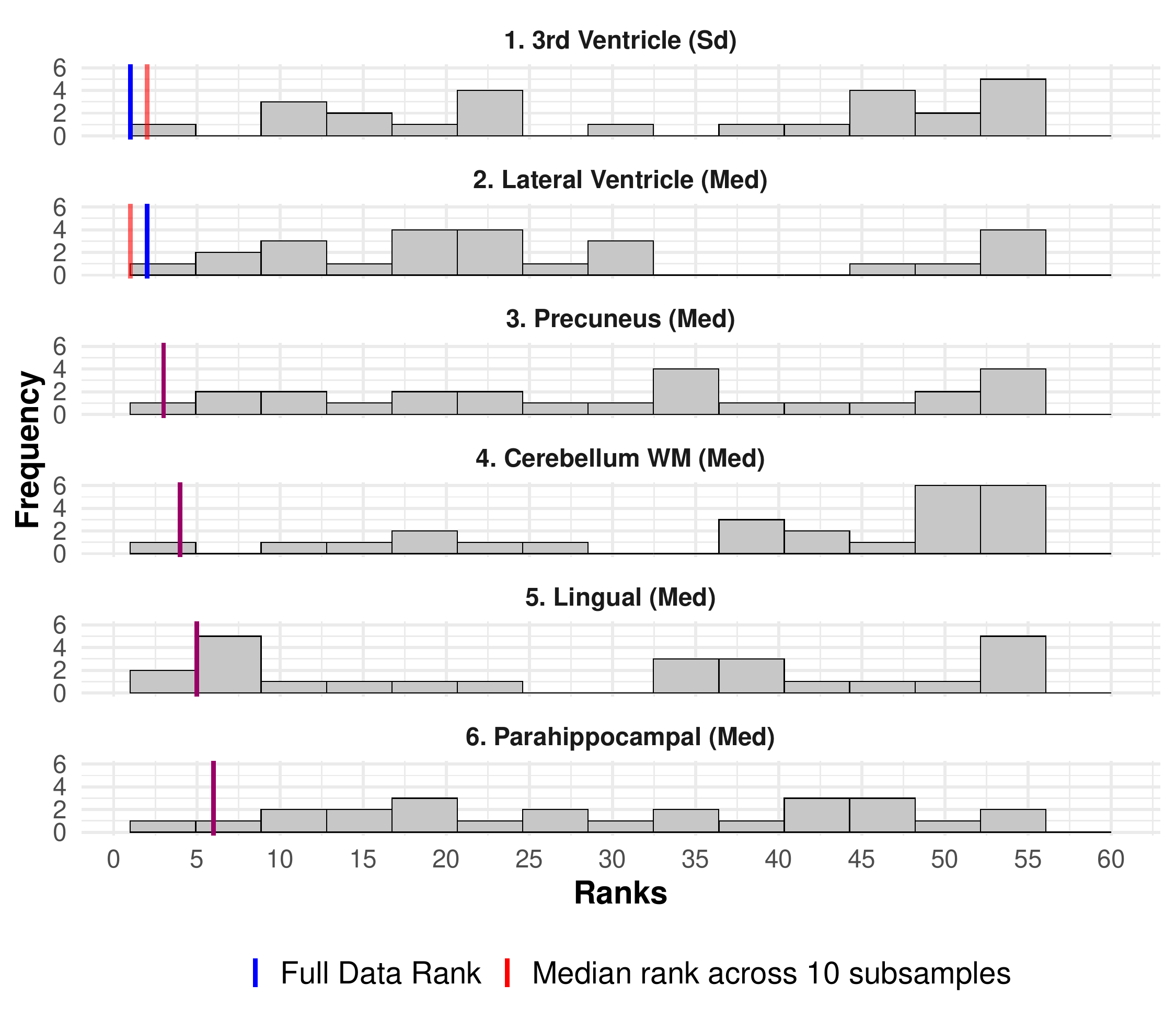}
    \caption{\smallrevision{LOCO-MP permutation test. We plot the distribution of the top six LOCO-MP features after 1) permuting their identities and 2) re-applying LOCO-MP to each permutation. The gray histograms show the rank distribution of each of the top LOCO-MP features after it is permuted. ``Med'' stands for voxel median and ``Sd'' stands for voxel standard deviation. The blue and red vertical lines represent the full data and median subsampled ranks (both unpermuted), respectively. These lines appear as a single purple line when the permuted and unpermuted ranks are identical.}}
    \label{fig:permuted-loco-ranks}
\end{figure}

\paragraph{Permutation test} We tested the individual importance of the top DBM features by permuting their values across all patients and evaluating the ranks. For each feature, we permuted its values 25 times across patients and re-applied LOCO-MP for each permutation to evaluate the ranking of the permuted feature. The permutation disrupts the connections between individual DBM features and the outcome and tests whether higher ranks are attributable to specific feature identities. If the ranks of the permuted features are close to those from the unpermuted data, it would imply that the observed importance of the feature is not driven by the uniqueness of the feature (and corresponding brain region) itself. Instead, the perceived feature importance may be due to random and/or other global patterns in the data.
The permuted ranks for each feature form a ``pseudo-null distribution" under the null hypothesis that the feature is uninformative (shown in Figure \ref{fig:permuted-loco-ranks}). This allows us to assess how often permuted ranks are as low or lower than the original rank. The results are shown in Figure \ref{fig:permuted-loco-ranks} -- note that ``Med'' stands for the voxel median feature and ``Sd'' stands for the voxel standard deviation feature. For the top features identified by LOCO-MP, the ranks from the unpermuted models consistently fall below most, if not all of the permuted ranks, yielding a low ``empirical p-value.'' This suggests that the feature identities are indeed meaningful and unlikely due to chance, highlighting their potential relevance as medically important regions for MS progression, discussed in the last paragraph of this section.

\begin{figure}[ht]
    \centering
    \includegraphics[width=1.05\linewidth]{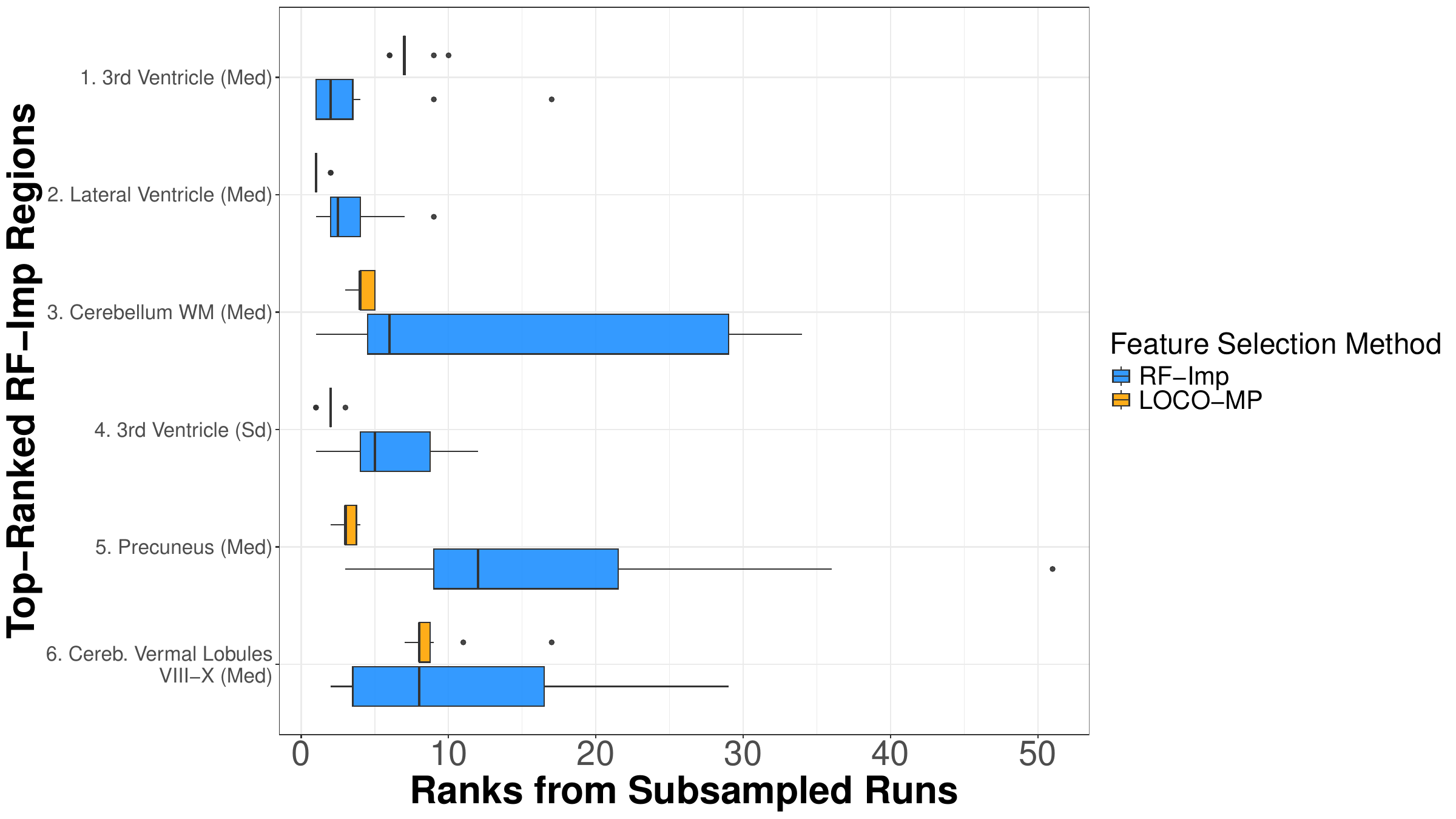}
    \caption{\smallrevision{Rank distribution for RF-Imp feature importance (blue) vs LOCO-MP (orange). The $y$-axis shows the top six RF-Imp features in terms of their feature importance score from an RSF model applied to the full data. The $x$-axis plots the ranks across different data subsamples. ``Med'' stands for voxel median and ``Sd'' stands for voxel standard deviation. The blue boxplots show their subsampled ranks on the $x$-axis. The orange boxplot above to each blue boxplot shows the rank distribution of that feature from LOCO-MP.}}
    \label{fig:rfgain-vs-loco}
\end{figure}

\paragraph{LOCO-MP vs Random forest-based scoring} 
\begin{revision}
    To assess the effectiveness of LOCO-MP, we benchmark it against the default feature importance scores produced by the random survival forest model which we refer to as ``RF-Imp.'' This comparison illustrates the differences between our LOCO-MP approach and the results one might obtain by directly applying standard survival modeling software without first considering the underlying assumptions. While such tools are easily accessible, they may not be tailored for the challenges of high-dimensional feature selection. 

    The feature importance scores in RF-Imp are generated as the average difference in out-of-bag prediction error across decision trees when the corresponding feature is permuted. In principle, the bagging and feature subsets for each decision tree share some similarities with minipatch ensembling. Incorporating shuffled feature values onto already-trained decision trees can create large erroneous changes to predictions that would otherwise have fine performance if retrained without the feature. This may introduce unnecessary instability to RF-Imp when compared with LOCO-MP. Following the same procedure as LOCO-MP, we first compute RF-Imp feature importance scores from the full dataset and then on ten subsamples. Figure \ref{fig:rfgain-vs-loco} shows the top six RF-Imp features based on the full data ranking. The blue boxplot shows the rank distribution across the ten subsamples, and the orange boxplot above it shows the corresponding LOCO-MP subsampled rank distribution for comparison. 
\end{revision}

Although there is alignment in the set of top regions captured between RF-Imp and LOCO-MP, the RF-Imp results alone demonstrate larger rank instability, especially beyond the third and lateral ventricles. In particular, the 5th and 6th ranked features differ between LOCO-MP and RF-Imp, but the top four are the same. It is clear that LOCO-MP has considerably lower variability in its rank distributions over the ten subsamples, establishing it as a more stable choice for feature selection. In contrast, the RF-Imp ranks have greater spread across subsamples, increasing the possibility of identifying spuriously meaningful features. \smallrevision{As an additional stability check, we replace random forest with gradient boosting machines (GBM) as the underlying survival algorithm in LOCO-MP and find that the subsample ranks of the top regions are concordant with RF, excluding the parahippocampal region (Table \ref{tab:gbm_loco}).} The greater stability of LOCO-MP allows us to be more confident that the selected features are genuinely meaningful and not due to noise or other artifacts of the data. \smallrevision{In future work, an ensembling of LOCO-MP with multiple survival models may yield an even more stable feature set.} 

Similar top features are identified when LOCO-MP is combined with survival gradient boosting machines (GBM) instead of random forest, highlighting that LOCO-MP can be relatively algorithm agnostic.

\paragraph{Clinical relevance of identified regions} 
In spite of the modestly sized LOCO-MP feature importance scores arising from the small cohort MRI data, the features and their ranks are stable. These features also align with brain regions that are clinically relevant in MS. The two identified ventricular regions, including the third and lateral ventricles, are often indicated in MS progression studies \cite{muller2013third,guenter2022predictive,simon1999longitudinal}.
\begin{revision}
    The consistency of ventricle-related findings across different studies emphasizes the robustness of these regions. Notably, these regions were also identified as significant in a recent causal study using the treatment arm of the same trial \cite{song2022deformation}. While this study focused on treatment effects, our use of the control arm to predict progression confirms the continued importance of the ventricular regions even in the absence of therapeutic intervention.
\end{revision}

The cerebellum is also frequently indicated in MS studies \cite{weier2015role,brouwer2024role}, which may play a role in impairment of postural control and balance in MS patients \cite{gera2020cerebellar}. The precuneus and lingual gyrus are found abnormally activated in the brains of MS patients when performing cognitive exercises, and this activation was later associated with increased mental fatigue and slower speed in task completion \cite{chen2020neural}. Finally, the parahippocampal gyrus (ranked 6th) has also been linked to varying activation patterns in MS patients compared to controls \cite{hulst2012functional}, and is also associated with much greater lesion activity compared to controls \cite{geurts2007extensive}.

\begin{revision}
    While the identified regions possess some degree of clinical relevance in previous MS studies, we emphasize that our study is exploratory in this area and does not intend to draw any clinical conclusions. Pathological development of MS can be diffusive across brain anatomy and functions \cite{COMPSTON20081502,eshaghi2018progression}. As a result, many brain regions can atrophy or be implicated by the disease. In order to inform future clinical decision-making, further validation on larger and more heterogeneous patient populations is required. However, our study provides a practical first step towards future scientific studies on certain brain regions.
\end{revision}

\begin{revision}
\subsection{Prediction improvements from selected regions under classical survival models} \label{sec:survmod-results}    
\end{revision}

We show that the predictive models comprising LOCO-MP identified features have stronger discriminative power than models using all DBM features.  We first present results from CCDP24, a canonical endpoint for measuring MS progression. We then discuss how the LOCO-MP selected features generalize well to alternative endpoints and compensate for the loss of key conventional predictors. 

\subsubsection{CCDP24 outcome} \label{sec:ccdp24-results}
\begin{figure}[h]
    \centering
    \includegraphics[width=0.98\linewidth]{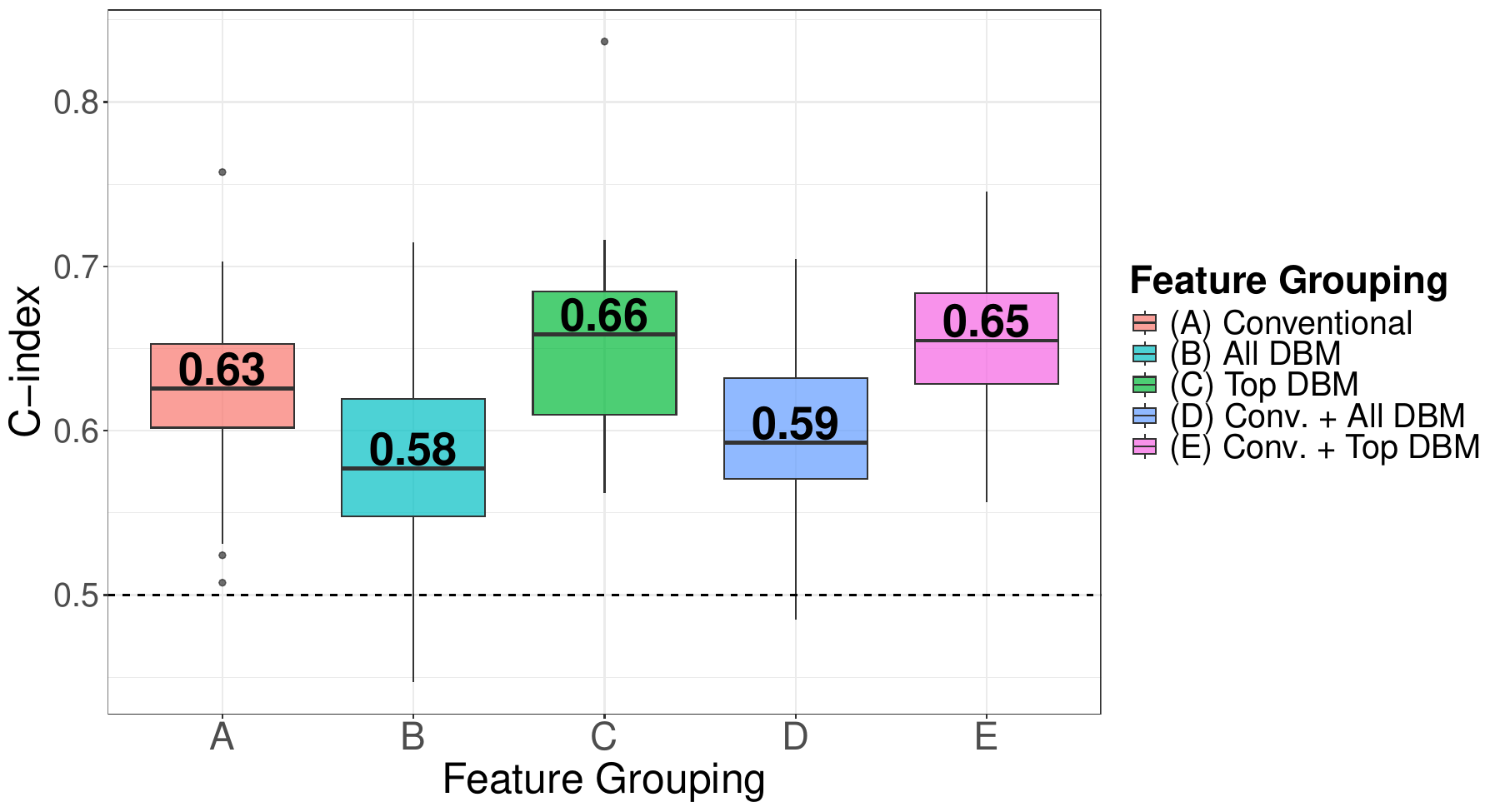}
    \caption{CCDP24 RSF model results. Boxplots show the distribution of random survival forest test-set C-indices across all cross-validation folds and repeats for outcome {CCDP24}. The $x$-axis consists of the feature groupings and the $y$-axis shows the test set C-indices. ``Top DBM" refers to the top six DBM features. The annotated C-index represents the median value for a specific feature grouping.}
    \label{fig:survmod-rf}
\end{figure}
Our RSF model results for the CCDP24 outcome are presented in Figure \ref{fig:survmod-rf}. Each boxplot shows the distribution of test-set C-indices for a specific feature grouping (discussed in Section \ref{sec:classical-survmod}) across all cross-validation folds and repeats. The median test-set C-index is annotated within each boxplot. Unsurprisingly, the ``All DBM" model has the lowest C-index across the feature groups, likely due to overfitting from the high dimensionality of using all 56 DBM features. A similar conclusion can be drawn from the ``Conventional + All DBM" model which includes even more features, showing how using an extensive set of MRI features without clinical context can lead to suboptimal predictive performance. These two models were slightly outperformed by the model using conventional features only, suggesting that the conventional features alone capture some degree of predictive value in CCDP24. 
In contrast, the two models that include the top features selected from DBM demonstrate the best prediction performance. Both C-indices are centered around 0.67, suggesting moderate predictive ability from the top DBM features. Adjusting for clinical information in the ``Conventional + All/Top DBM'' models did not appear to affect performance.

\begin{figure}[h]
    \centering
    \includegraphics[width=0.95\linewidth]{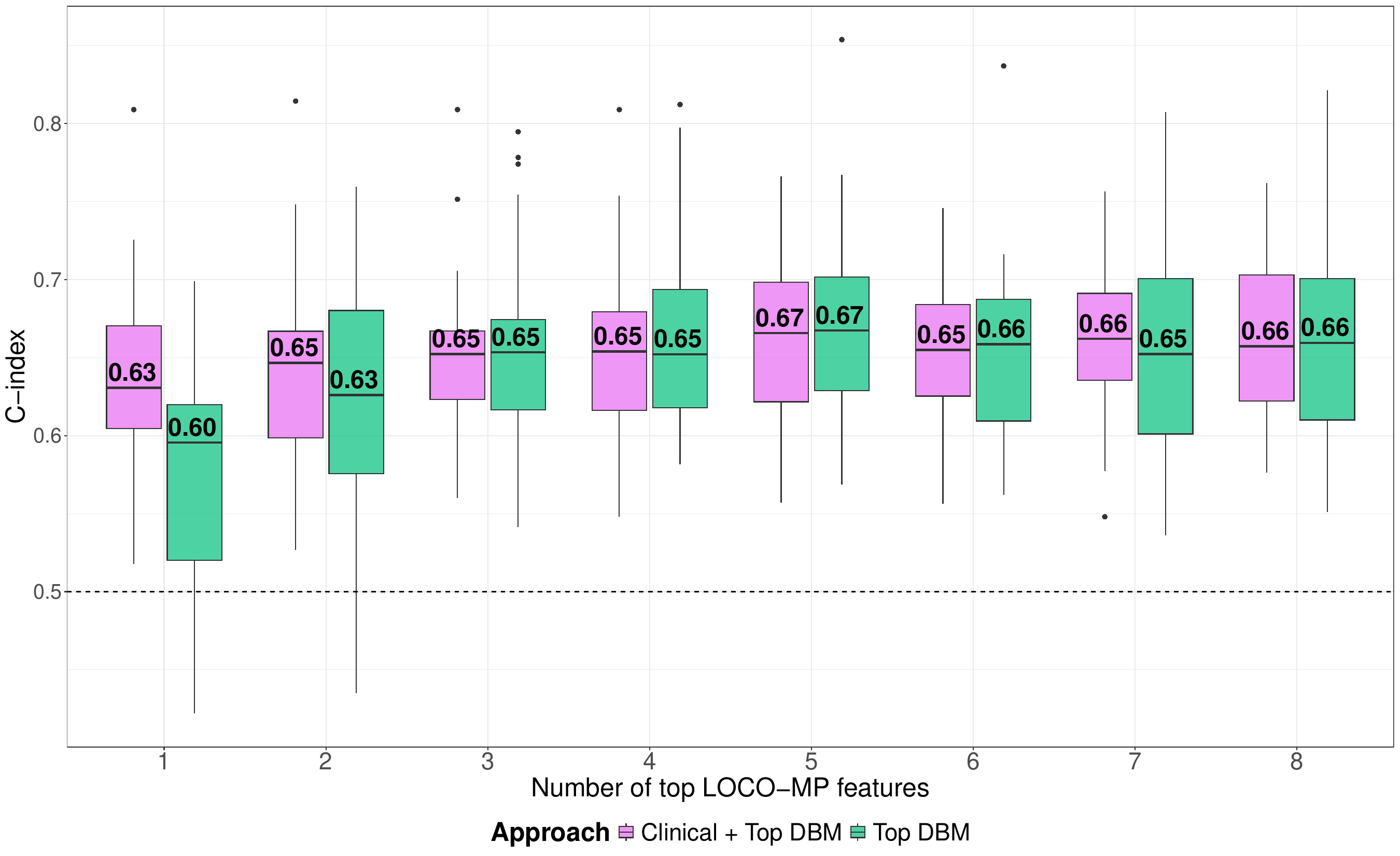}
    \caption{CCDP24 model stability. Boxplots show the distribution of CCDP24 test-set C-indices for the Conventional + Top DBM and Top DBM-only RSF models, plotted across varying numbers of top LOCO-MP features. The $x$-axis represents the number of top LOCO-MP features selected and the $y$-axis is the C-index.}
    \label{fig:survrf-ccdp-multitop-loco}
\end{figure}

For simplicity, the ``Top DBM'' models (Top DBM, Conventional + Top DBM) shown in Figure \ref{fig:survmod-rf} are fit with the top six LOCO-MP features. However, the results from these two feature groupings remained stable with respect to varying numbers of LOCO-MP-selected features. Figure \ref{fig:survrf-ccdp-multitop-loco} plots the distribution of test-set C-indices from these two feature groupings across the top-ranked DBM features, ranging from 1 to 8. Performance remains stable across different numbers of features, suggesting that the model is robust to how many top features are used -- the precise rankings between features and the specific number used for modeling is less important than identifying a broad set of features that can be investigated for clinical relevance. Although these two feature groupings are more investigative, researchers can be confident that the results are not overly sensitive to the specific subset of features. \smallrevision{To assess generalizability of the Top DBM features, Section \ref{sec:s25fw-results} applies this feature set to an alternative progression outcome, and Section~\ref{sec:generalizability} applies this feature set to a separate control cohort.}

Several conclusions can be drawn from our results. First, many of the DBM features contain overlapping information that ultimately results in model overfitting, as evidenced by the two ``All DBM" model boxplots. This underscores the value of LOCO-MP, which can sift through highly correlated groups of features to identify the most prognostic ones and improve prediction performance. Moreover, the predictive strength of the DBM features and the clinical relevance of the corresponding brain regions reinforces the validity of these findings, demonstrating that LOCO-MP identifies DBM features with genuine predictive value.


\subsubsection{25-foot walking test outcome} \label{sec:s25fw-results}

\begin{figure}[ht]
    \centering
    \includegraphics[width=0.95\linewidth]{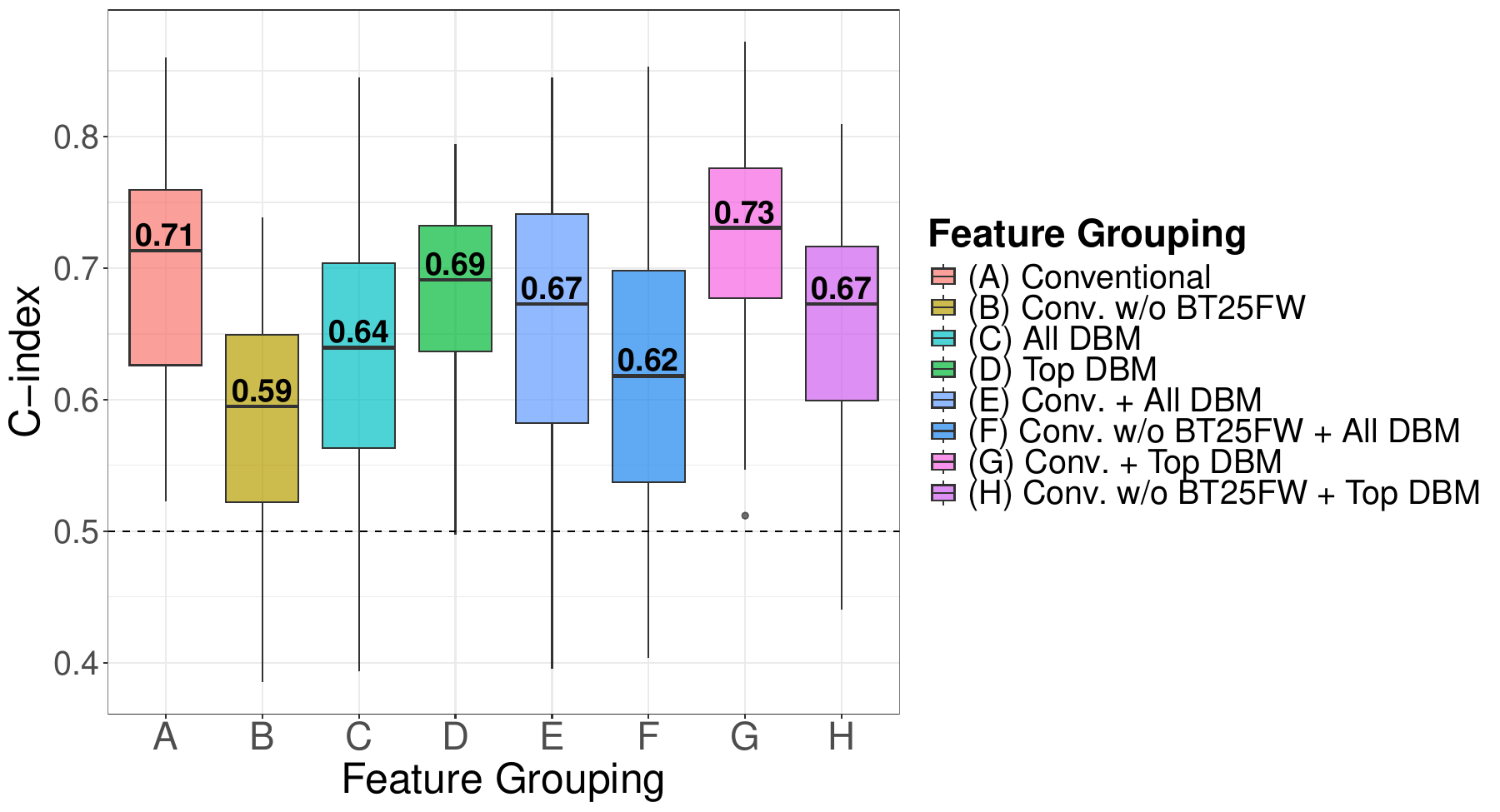}
    \caption{S25FW RSF model results. Boxplots show the distribution of random survival forest test-set C-indices across all cross-validation folds and repeats for outcome {S25FW}. The $x$-axis consists of the feature groupings and the $y$-axis shows the test set C-indices. ``Top DBM" refers to the top six DBM features. The annotated C-index represents the median value for a specific feature grouping. We also include results after omitting the highly prognostic conventional variable BT25FW.}
    \label{fig:survmod-rf-25fw}
\end{figure}
\begin{revision}
    Despite variations in performance from across different feature sets, the CCDP24 C-indices are modest for even the strongest feature groups. This may arise from various factors, including inherent limitations in the outcome definition which impacts the model's ability to identify meaningful signal -- CCDP24 is characterized by marked changes in multiple events that are sustained for 24 weeks \cite{montalban2017ocrelizumab,krishnan2023multi}. Although combining multiple events reduces the censoring rate, it comes at the cost of potentially introducing subjective measurements into the analysis: EDSS scoring can vary across studies, and the definition of EDSS-based progression highly depends on the study at hand \cite{krajnc2021measuring}. Therefore, including information from noisy outcome measurements could potentially worsen performance and we hypothesize this may be the case with CCDP24. To combat this, recent literature has pointed to the timed 25-foot walking test as an alternative measure of disability progression that may be more sensitive in detecting genuine disease worsening \cite{krajnc2021measuring}.
\end{revision}
This test measures the time for a patient to walk 25 feet without any assistive devices. Progression is defined as a 20\% increase in 25-foot walking time that is sustained for 24 weeks (S25FW). Previous studies have demonstrated significantly worse S25FW times in MS patients \cite{sikes2020quantitative} and S25FW has previously been shown to be strongly correlated with other forms of MS progression \cite{koch2020clinical}. 

\begin{figure}[ht]
    \centering
    \includegraphics[width=0.95\linewidth]{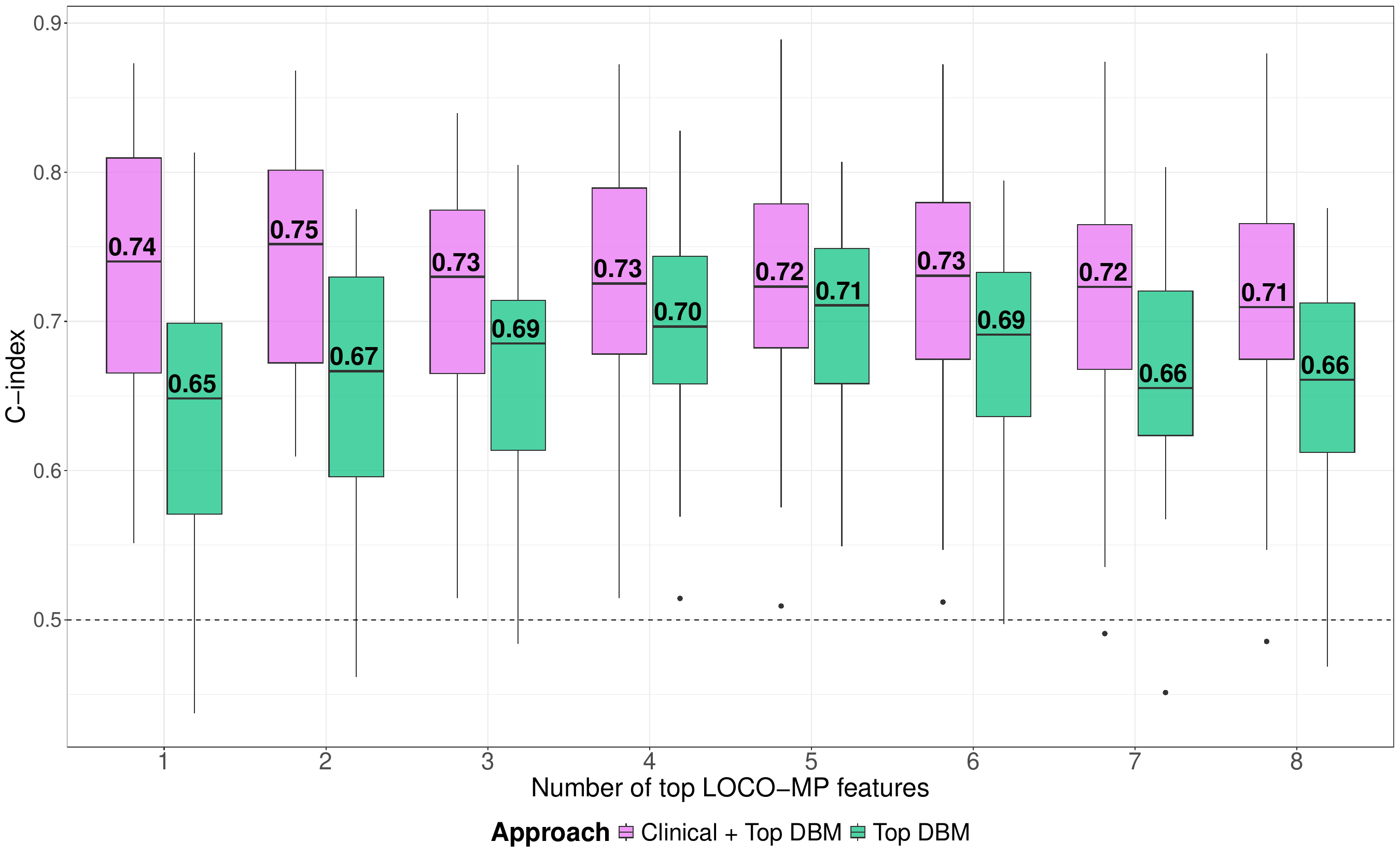}
    \caption{S25FW model stability. Boxplots show the distribution of S25FW test-set C-indices for the Conventional + Top DBM and Top DBM-only RSF models, plotted across varying numbers of top LOCO-MP features. The $x$-axis represents the number of top LOCO-MP features selected and the $y$-axis is the C-index.}
    \label{fig:survrf-25fw-multitop-loco}
\end{figure}

This outcome has a higher censoring rate compared to CCDP24 (88\% compared to 77\%), but resulted in higher C-indices across all feature groupings, shown in Figure \ref{fig:survmod-rf-25fw}. Similar to CCDP24, a model using all DBM features leads to overfitting. However, the conventional feature model demonstrates a 12.7\% increase in median C-index compared to CCDP24 (0.63 to 0.71), and the ``Conventional + Top DBM'' and ``Top DBM-only'' models also see increases in median C-index. These increases suggest that DBM information is more effective in predicting S25FW-based progression. This also supports previously stated claims about the improved sensitivity of S25FW, illustrating the perils of using a noisy outcome measurement like CCDP24. The C-index for S25FW also remained robust to different numbers of top DBM features, as shown in Figure \ref{fig:survrf-25fw-multitop-loco}.

The strong performance in the conventional model for S25FW is largely attributed to the baseline 25-foot walking test feature (BT25W, see Section \ref{sec:data}) which is highly predictive of S25FW progression. The C-indices decrease after omitting BT25W as a conventional covariate in our models -- the ``Conventional-only'' and ``Conventional + All DBM'' models achieve a median C-index of 0.59 and 0.62, respectively, as shown in Figure \ref{fig:survmod-rf-25fw}. These performances are comparable to their CCDP24 counterparts. While omitting BT25FW results in worse performance across all feature groupings with conventional features, the ``Conventional + Top DBM'' model performs fairly robustly even after omitting BT25FW, achieving a median C-index of 0.67. This demonstrates that the top DBM features may be able to compensate for the loss of key conventional predictors and shows how LOCO-MP can identify robust and generalizable features.
 
We also note that the two ``Top DBM" models in the S25FW analysis use LOCO-MP features identified from the CCDP24 outcome -- the algorithm encountered challenges when applied to S25FW due to its high censoring rate, which limited the number of informative minipatches (many minipatches consisted entirely of censored patients). Despite this, the increased performance in the two ``Top DBM" S25FW models demonstrates the reliability of LOCO-MP in identifying a stable feature set that captures fundamental aspects of disease progression that are robust across different outcome definitions. \smallrevision{\ref{sec:generalizability} further demonstrates that these identified DBM improve S25FW prediction in a separate validation cohort. These strong generalizations are} further evidence that the top features may be genuine markers of disease worsening, enhancing its potential for broader clinical applications and underscoring the value of DBM as a regional identification framework.

While this new outcome allowed for more precise patient risk discrimination, the variance of the C-index estimates was higher than that of CCDP24. This is due to the higher censoring rate in S25FW which, by reducing the number of informative events, lowers the effective sample size. This generally leads to increased variance in the estimates because there is less observed data available to accurately capture the underlying risk. Even though S25FW provides clearer risk stratification, the higher variation in the C-index reflects the statistical challenges introduced by a lower effective sample size. In contrast, the CCDP24 outcome has lower variance but lower C-indices as well.

\begin{revision}
\subsection{Prediction improvements from selected regions under convolutional neural networks} \label{sec:cnn-results}    
\end{revision}

\begin{figure}[h]
    \centering    
    \includegraphics[width=0.85\linewidth]{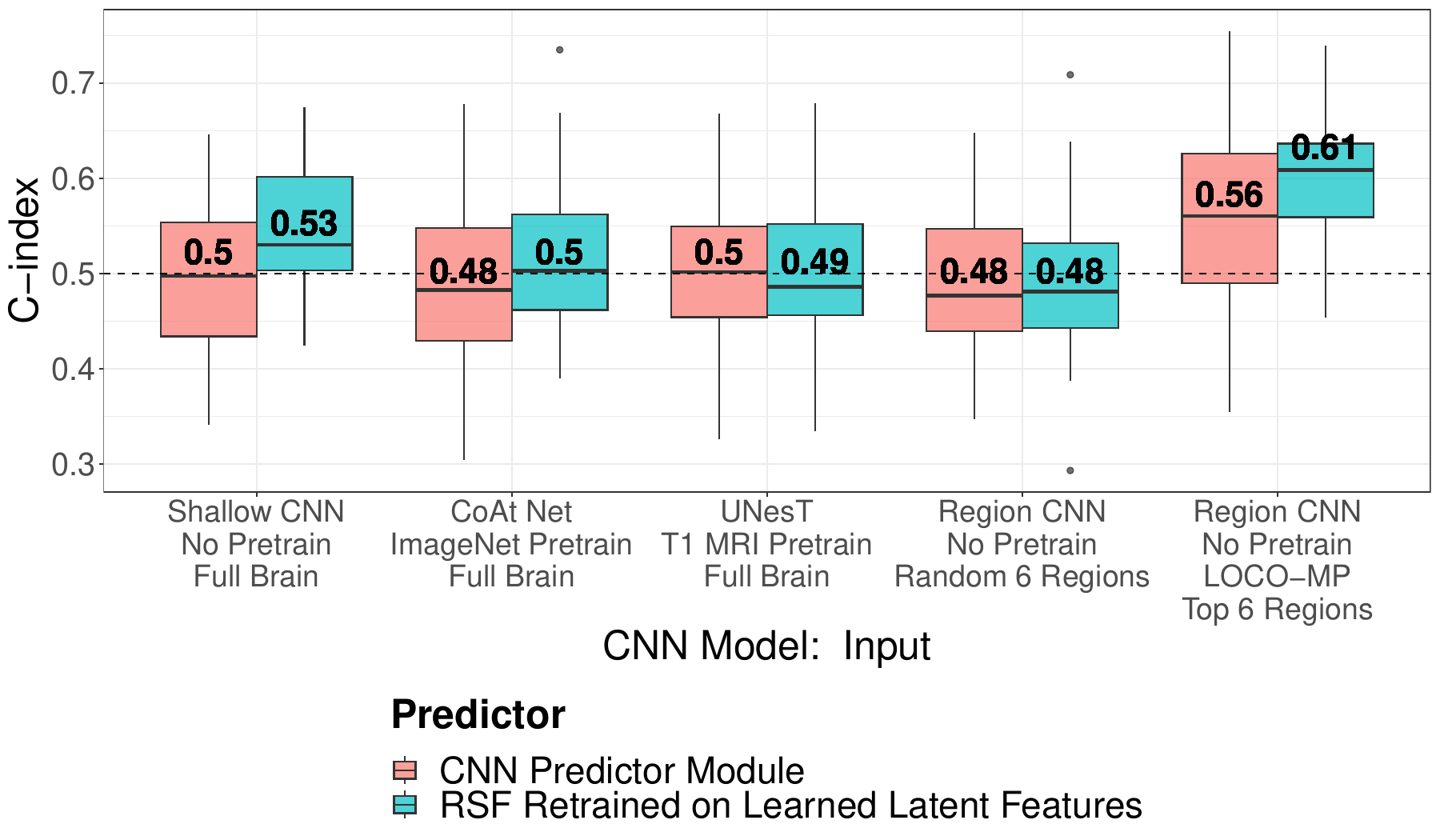}
    \caption{CNN model performance for CCDP24. Boxplots show the distribution of test C-indices for each of the tested CNN models. Results are presented both for the trained CNN predictor module (CNN Predictor Module), as well as an RSF trained on the extracted feature vector of each trained CNN (RF Retrained on Extracted Features). The $x$-axis labels first give the CNN model architecture used, followed by the inputs used. Specifically, full brain models are compared with \textit{Region CNN}, using either the top 6 LOCO-MP regions or random 6 regions (excluding 6 LOCO-MP regions) with each fold.}
    \label{fig:cnn_cindex_both}
\end{figure}


We compare full brain CNNs against \textit{Region CNN} using only the top 6 identified LOCO-MP regions to assess the extent to which an unsupervised deep learning feature extraction improves prediction performance. 
Figure \ref{fig:cnn_cindex_both} presents CNN model test C-indices over 30 80\%/20\% data splits for predicting CCDP24.  The evaluated models include the three baseline full brain models (Shallow, CoAtNet UNesT), \textit{Region CNN} trained on the top 6 LOCO-MP regions, and \textit{Region CNN} trained on 6 randomly selected regions (excluding the top 6 LOCO-MP regions). Each of these models are described in Section \ref{sec:3d-cnn}. Red boxplots show the C-index distribution from the predictions of the CNN predictor module. The learned latent features from each CNN (i.e., the concatenated green bar in Figure \ref{fig:fb_cnn_diag}) were also extracted and retrained using a conventional RSF survival model (without conventional features), enabling the continuous survival times to be used as the outcome. Teal boxplots show the C-index distribution of the RSF models trained on these extracted CNN feature vectors.

These results underscore the importance of subsetting the DBM input to the top 6 LOCO-MP regions in order to extract signal from a CNN. Each of the full brain CNN models struggle to achieve nonrandom performance (C-index $\geq 0.5$). The shallow full brain CNN features perform modestly better than the complex CoAtNet and UNesT architectures. This suggests that, for this data and cohort size, neither an attention mechanism nor a very deep architecture are able to adequately predict CCDP24. Additionally, CoAtNet was pretrained using ImageNet \cite{imagenet} and UNesT was pretrained using raw T1-w MRI tensors, indicating that the DBM data may reflect critically different patterns than the raw MRI. \textit{Region CNN} fares quite poorly when trained on 6 random non-LOCO-MP regions. However, it extracts strong predictive signal when LOCO-MP features are used. The LOCO-MP features were more helpful in the RSF model, achieving a median C-index of 0.61. \textit{Region CNN} with the top 6 LOCO-MP regions performs comparably on the S25FW outcome (median predictor module C-Index: $0.57$; median RSF retrained C-Index: $0.61$). Ultimately, even the best CNN fares moderately worse than the corresponding classical survival model in Section \ref{sec:survmod-results}, highlighting the challenge of learning complex models for the DBM data on this cohort size. With more data, \textit{Region CNN} (and CNNs generally) may be able to extract more expressive features from the T1-w MRI data.



\begin{revision}
\section{Discussion} \label{sec:modeling-discussion}
This study develops a data analysis pipeline that balances stability, predictive performance, and domain-level interpretability to improve MS prediction models with DBM-based features. We demonstrate how LOCO-MP identifies clinically relevant features in a stable manner, outperforming other feature importance approaches. Our sample size of 350 is modest for complex prediction tasks such as MS progression prediction from neuroimaging data. While sample size can influence the stability and generalizability of model outputs, we mitigated these concerns by conducting extensive stability checks discussed in Section \ref{sec:feat-imp} and varying the number of top DBM features used for modeling (Figures \ref{fig:survrf-ccdp-multitop-loco} and \ref{fig:survrf-25fw-multitop-loco}). The selected LOCO-MP features generalized well to a separate validation cohort (\ref{sec:generalizability}). We note that our analysis pipeline is scalable and well-suited to larger datasets as they become available.

Regarding prediction performance, we observe that the test set C-indices for the strongest feature groupings primarily range between 0.65-0.75. We note that our prediction task focuses on 24-week progression, a relatively short time frame for a chronic illness such as MS. Predicting over such brief periods is inherently challenging, as disease progression tends to be subtler and more variable, limiting the ability of models to capture robust patterns. This limitation likely prevents the observed C-index from being higher. In fact, several other RRMS patient studies that predict short-term progression with similar sample sizes report comparable C-indices or AUROC values\footnote{The C-index is analogous to the AUROC in binary prediction tasks with survival data.} (between 0.48 and 0.73) when using a similarly-defined outcome with different types of brain data \cite{pellegrini2020predicting,branco2023survival,guazzo2023baseline,andorra2024predicting}.
Another study reported slightly higher C-indices ($0.76$) when integrating genotypic factors with clinical variables \cite{fuh2021developing} -- the use of non-MRI based features, such as those based on genetic or gut microbiome information, has proven to yield strong predictions and is a potential future
direction for MS prediction \cite{hossain2022role}. In addition, Pellegrini et al. (2020) observed that C-indices did not exceed 0.65 with conventional survival models, despite using a larger cohort \cite{pellegrini2020predicting}. 

The endpoints themselves also have limitations. As mentioned in Section \ref{sec:s25fw-results}, CCDP24 is derived by combining the scores of EDSS, 9HPT, and T25FW.  While each of the three clinical assessments has its own limitations \cite{cohen2021should,s25fw_paper,solaro2019clinical}, EDSS is more subjective and less sensitive to disease-related changes \cite{cohen2021should}. S25FW is a quantitative measure of lower limb movement function. Because it only captures one specific function, its two-year progression rate experiences more censoring. Progression on this metric regardless has a high correspondence with overall RRMS progression \cite{s25fw_paper}. It remains an area of future work to find a clinical progression measure that optimally balances sensitivity and censoring rates.

We additionally explored the performance of 3D CNN models using the full brain versus the LOCO-MP top regions. Though the 3D CNN models substantially improved when constraining the DBM input to the LOCO-MP top DBM regions, their performance still lagged behind conventional modeling using summary statistics of these regions. State of the art 3D CNN architectures did not change this story; the small sample size could be the main limiting factor for deep learning performance. In the future, we are interested in a deeper exploration of small sample size strategies for CNNs, such as DBM pretraining with self-supervised learning (SSL) on DBM-processed T1 MRI data, model ensembling, domain-tailored attention, and few-shot learning techniques.


Ultimately, our results suggest that DBM uniquely captures subtle volumetric abnormalities across the brain and provides orthogonal information to the conventional features, contributing to MS progression in a way that is stable and clinically meaningful across multiple definitions of progression. Future work could integrate DBM with other MRI modalities used for lesion localization. Additionally, our study was restricted solely to patients with relapsing MS. Future work could utilize DBM to differentiate between different subtypes of MS or predict the conversion of relapsing MS to secondary progressive MS \cite{preziosa2022slowly} when those subtypes are available. Our analysis pipeline is an initial starting point for more advanced data-driven studies that may provide better scientific and clinical insights about how MS impacts the brain.

\end{revision}

\section*{Acknowledgments and funding}
Andy Shen is partially supported by the National Science Foundation (NSF) Graduate Research Fellowship under Grant No. 2146752. Any opinion, findings, and conclusions or recommendations expressed in this material are those of the authors(s) and do not necessarily reflect the views of the NSF. This project was supported by Genentech/Roche.

\section*{Declaration of competing interest}
Richard A.D. Carano and Zhuang Song are employees of Genentech/Roche and are stockholders of Roche.

\section*{Data availability statement}
The fully anonymized, individual patient raw data including clinical and MRI data of the OPERA trials are made available through the International Progressive MS Alliance (\href{www.progressivemsalliance.org}{www.progressivemsalliance.org}).


\bibliographystyle{elsarticle-num} 
\bibliography{paper_refs}

\newpage
\appendix



\begin{revision}
\section{Details of LOCO-MP for survival analysis} \label{sec:loco-details}
Following the notation of Gan et al. (2023) \cite{gan2023modelagnosticconfidenceintervalsfeature}, we denote the $N \times M$ data matrix as $X$ and the outcome as $Y$. Let $(X_i, Y_i)$ denote the feature-outcome pair for individual $i$, where $X_i$ is a $M$-dimensional vector and $Y_i = (T_i, C_i)$. For every minipatch $k = 1 \dots K$, we sample $n$ observations $I_k$ and $m$ features $F_k$ from $X$. Denote this minipatch as $\XIkFk$. 
A prediction model $\mukhat$ is then trained on $\XIkFk$ and predictions are generated for the remaining observations $\XminusIkFk$, which includes features $F_k$ and observations outside $I_k$. Repeating this over all $K$ minipatches, we compute the individual feature occlusion score for feature $j$ and observation $i$:
\begin{align*}
\label{eqn:loco-error-i}
    \hat{\Delta}_{ij} = \text{Error}(Y_i, \hat{\mu}_{-i}^{-j}(X_i)) - \text{Error}\left(Y_i, \hat{\mu}_{-i}(X_i)\right),
\end{align*}
where 
\begin{align*}
    \hat{\mu}_{-i}(X_i) &= \frac{1}{\sum_{k=1}^{K} \mathbbm{1}(i \notin I_k)} \sum_{k=1}^{K} \mathbbm{1}(i \notin I_k) \hat{\mu}_k(X_i), \text{ and} \\
    \hat{\mu}_{-i}^{-j}(X_i) &= \frac{1}{\sum_{k=1}^{K} \mathbbm{1}(i \notin I_k) \mathbbm{1}(j \notin F_k)} \sum_{k=1}^{K} \mathbbm{1}(i \notin I_k) \mathbbm{1}(j \notin F_k) \hat{\mu}_k(X_i),
\end{align*}
and $\text{Error}()$ refers to any observation-level prediction error function, in our case the discrete hazard loss discussed in the following paragraph. The individual $\hat{\Delta}_{ij}$ values are then averaged, giving us the feature occlusion score for feature $j$:
\begin{equation}
    \label{eqn:loco-avg-error}
    \Bar{\Delta}_j =\frac{1}{N}\sum_{i=1}^N \hat{\Delta}_{ij}.
\end{equation}
$\hat{\Delta}_{ij}$ is the difference in error for observation $i$ when feature $j$ is omitted. Concretely, this can be thought of as the change in prediction capability for patient $i$ when a specific DBM feature is omitted. Larger, positive values of $\hat{\Delta}_{ij}$ indicate greater feature importance since the model performs worse (has larger error) when feature $j$ is excluded. 

The prediction error for survival analysis models is commonly measured as the reciprocal of the overall concordance index (C-index) of risk scores between pairs of observations. However, this metric is not an aggregation of observation-level errors, preventing its direct usage in the LOCO-MP framework. To address this, we utilize discrete hazard loss functions used in deep learning survival analysis models \cite{zadeh2020bias, vale2021long, shao2023hvtsurv}. Specifically, we evenly divide the survival time scale of the study into $d$ intervals: $[t_0, t_1), \hdots, [t_{d-1}, t_d)$, where $t_d$ is the end of the study period. The patient's event time is now denoted $T_i = q \; \text{if and only if} \; T_i \in [t_q, t_{q+1})$.

Given this setup, we may define a patient's conditional hazard probability as:
\begin{align*}
    h(q \mid X_i) = \mathbb{P}(T_i=q \mid T_i \geq q, X_i).
\end{align*}

This patient's survival probability follows as the probability of surviving until the end of the current interval or longer:
\begin{align*}
    S(q \mid X_i) = \mathbb{P}(T_i > q \mid X_i) = \prod_{s=1}^q (1 - h(s\mid X_i)).
\end{align*}

From these definitions, we can compute an observation-level likelihood.  For uncensored patients, this is the product of the conditional hazard at their event time and the survival function of the prior time period:
\begin{align*}
    \ell_{U}(T_i =q) = h(q \mid X_i)~S(q-1 \mid X_i).
\end{align*}

For patients that are right-censored at time period $q$, their likelihood is simply the survival function at time period $q$:
\begin{align*}
    \ell_{C}(T_i>q) = S(q \mid X_i).
\end{align*}

Altogether, given the censoring status $C_i$ for each patient, the observation level model error for LOCO-MP at time $T_i = q$ is defined as the negative log-likelihood of the discretized observation:
\begin{align}
    \text{Error}((T_i, C_i), \hat{\mu}_{-i}^{-j}) =
    \begin{cases}
        -\log(h(q \mid X_i)) - \log(S(q-1 \mid X_i));& \; C_i=0 \\
        -\log(S(q \mid X_i));& \; C_i=1
    \end{cases}
    \label{eqn:log_lik_loss}
\end{align}

Here, $\hat{\mu}_{-i}^{-j}$ encapsulates the collection of conditional hazards that are computed with the underlying minipatch model. Assuming that the patient conditional hazard probabilities are not exactly 0 or 1 over the study period, this loss framework satisfies the necessary assumptions for the theoretical guarantees of LOCO-MP \cite{gan2023modelagnosticconfidenceintervalsfeature}.

LOCO-MP is advantageous over other feature selection techniques due to its ability to provide asymptotic inferential guarantees  \cite{gan2023modelagnosticconfidenceintervalsfeature}. Moreover, LOCO-MP can also account for dependencies across features, a common issue in high-dimensional settings like ours --  Gan et al. (2023) \cite{gan2023modelagnosticconfidenceintervalsfeature} discuss that, by generating randomly subsampled features across minipatches, LOCO-MP ensures that the predictive value of each feature is not being diminished by other strongly correlated features, since groups of strongly correlated features will not always appear in the same minipatch. The full LOCO-MP procedure and its theoretical guarantees are discussed in Gan et al. (2023) \cite{gan2023modelagnosticconfidenceintervalsfeature}.
\end{revision}

\section{Additional LOCO-MP and feature selection results}

\subsection{Full LOCO-MP results} \label{sec:loco-full-tables}
\begin{figure}[ht]
    \centering
    \includegraphics[width=1.1\linewidth]{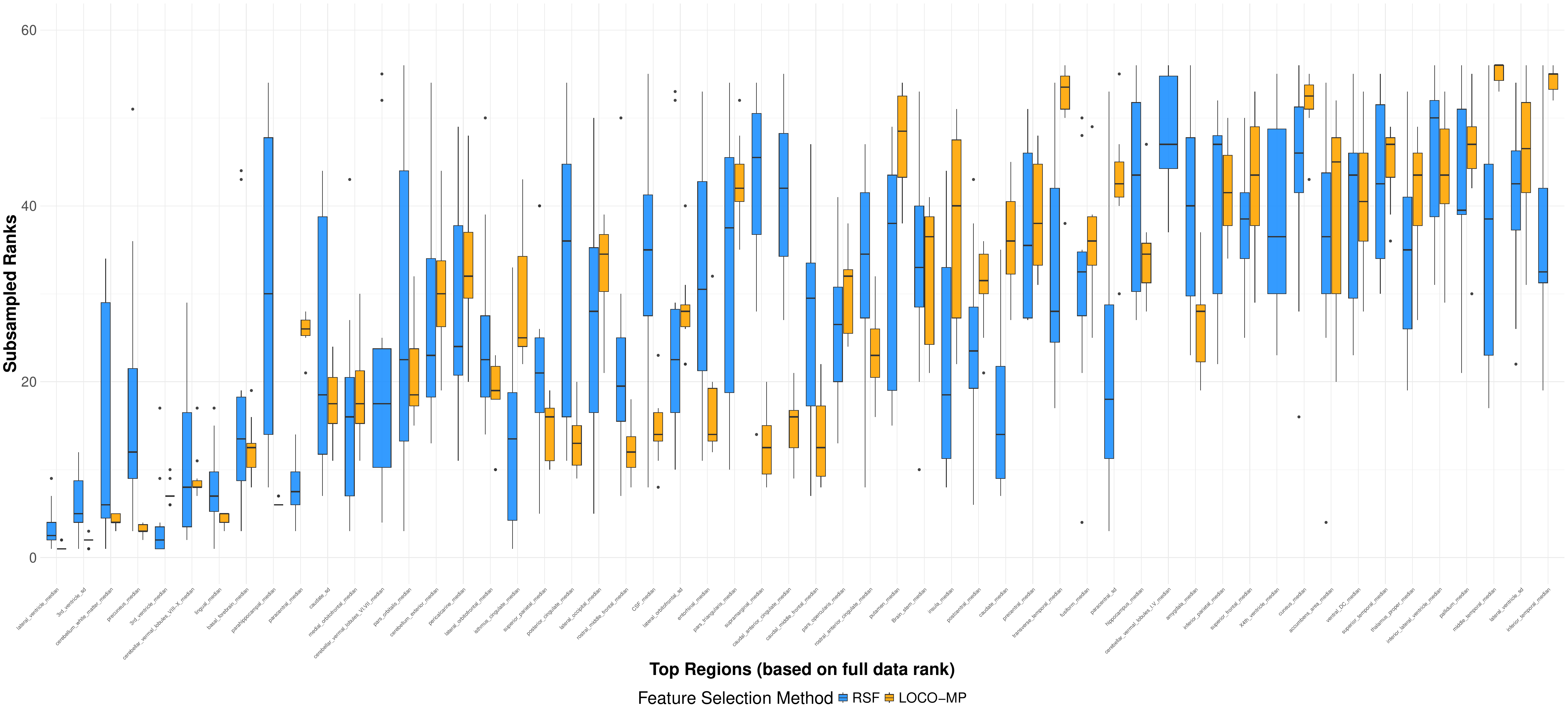}
    \caption{Rank distribution for RSF feature importance (blue) vs LOCO-MP (orange) for all DBM features. The $x$-axis shows the features (from highest to lowest rank) in terms of their feature importance score from an RSF model applied to the full data, and the blue boxplots show their subsampled ranks on the $y$-axis. The orange boxplot next to each blue boxplot shows the rank distribution of that feature from LOCO-MP.}
    \label{fig:rfgain-all-feats}
\end{figure}

Figure \ref{fig:rfgain-all-feats} shows boxplots for all 56 DBM features, similar to Figure \ref{fig:rfgain-vs-loco} in the main text. Blue boxplots show the rank distribution from RF-Imp while orange boxplots show the rank distribution from LOCO-MP.

Table \ref{tab:all_regions_loco} shows the LOCO-MP feature importance scores for all 56 DBM features.

\begin{table}[ht]
\scriptsize
\centering
\begin{tabular}{rlcc|cc}
\toprule
& \textbf{Region} & \multicolumn{2}{c|}{\textbf{Full Data}} & \multicolumn{2}{c}{\textbf{Subsamples}} \\ 
\cmidrule(lr){3-4} \cmidrule(lr){5-6}
& & \textbf{Rank} & $\Bar{\Delta}_j$ & \textbf{Median Rank} & \textbf{Median $\Bar{\Delta}_j$} \\ 
\midrule
& 3rd Ventricle SD & 1 & 0.00104 & 2 & 0.00090 \\ 
& Lateral Ventricle Median & 2 & 0.00101 & 1 & 0.00095 \\ 
& Precuneus Median & 3 & 0.00088 & 3 & 0.00078 \\ 
& Cerebellum White Matter Median & 4 & 0.00056 & 4 & 0.00064 \\ 
& Lingual Median & 5 & 0.00051 & 5 & 0.00061 \\ 
& Parahippocampal Median & 6 & 0.00051 & 6 & 0.00042 \\ 
& Cerebellar Vermal Lobules VIII-X Median & 7 & 0.00029 & 8 & 0.00027 \\ 
& 3rd Ventricle Median & 8 & 0.00028 & 7 & 0.00037 \\ 
& Basal Forebrain Median & 9 & 0.00026 & 12.5 & 0.00016 \\ 
& Caudal Anterior Cingulate Median & 10 & 0.00023 & 16 & 0.00014 \\ 
& Caudal Middle Frontal Median & 11 & 0.00017 & 12.5 & 0.00019 \\ 
& Supramarginal Median & 12 & 0.00016 & 12.5 & 0.00017 \\ 
& Rostral Middle Frontal Median & 13 & 0.00016 & 12 & 0.00019 \\ 
& CSF Median & 14 & 0.00015 & 14 & 0.00015 \\ 
& Lateral Orbitofrontal Median & 15 & 0.00014 & 19 & 0.00008 \\ 
& Posterior Cingulate Median & 16 & 0.00014 & 13 & 0.00017 \\ 
& Pars Orbitalis Median & 17 & 0.00012 & 18.5 & 0.00008 \\ 
& Caudate SD & 18 & 0.00010 & 17.5 & 0.00009 \\ 
& Medial Orbitofrontal Median & 19 & 0.00005 & 17.5 & 0.00008 \\ 
& Paracentral Median & 20 & 0.00004 & 26 & -0.00002 \\ 
& Pericalcarine Median & 21 & 0.00003 & 32 & -0.00010 \\ 
& Superior Parietal Median & 22 & 0.00002 & 16 & 0.00014 \\ 
& Cerebellum Exterior Median & 23 & -0.00000 & 30 & -0.00007 \\ 
& Entorhinal Median & 24 & -0.00002 & 14 & 0.00014 \\ 
& Amygdala Median & 25 & -0.00002 & 28 & -0.00003 \\ 
& Brain Stem Median & 26 & -0.00003 & 36.5 & -0.00014 \\ 
& Rostral Anterior Cingulate Median & 27 & -0.00003 & 23 & 0.00001 \\ 
& Lateral Occipital Median & 28 & -0.00005 & 34.5 & -0.00011 \\ 
& Isthmus Cingulate Median & 29 & -0.00005 & 25 & -0.00002 \\ 
& Pars Triangularis Median & 30 & -0.00006 & 42 & -0.00021 \\ 
& Superior Frontal Median & 31 & -0.00007 & 43.5 & -0.00021 \\ 
& Pars Opercularis Median & 32 & -0.00008 & 32 & -0.00007 \\ 
& Lateral Orbitofrontal SD & 33 & -0.00010 & 28 & -0.00005 \\ 
& Hippocampus Median & 34 & -0.00010 & 34.5 & -0.00011 \\ 
& Cerebellar Vermal Lobules I-V Median & 35 & -0.00010 & 25.5 & -0.00003 \\ 
& Caudate Median & 36 & -0.00012 & 36 & -0.00012 \\ 
& Fusiform Median & 37 & -0.00012 & 36 & -0.00015 \\ 
& Accumbens Area Median & 38 & -0.00013 & 45 & -0.00022 \\ 
& Insula Median & 39 & -0.00019 & 40 & -0.00020 \\ 
& Inferior Parietal Median & 40 & -0.00019 & 41.5 & -0.00019 \\ 
& Postcentral Median & 41 & -0.00019 & 31.5 & -0.00010 \\ 
& Pallidum Median & 42 & -0.00020 & 47 & -0.00023 \\ 
& Ventral DC Median & 43 & -0.00020 & 40.5 & -0.00018 \\ 
& Precentral Median & 44 & -0.00023 & 38 & -0.00016 \\ 
& Transverse Temporal Median & 45 & -0.00023 & 53.5 & -0.00035 \\ 
& Cerebellar Vermal Lobules VI-VII Median & 46 & -0.00024 & 49 & -0.00026 \\ 
& Putamen Median & 47 & -0.00025 & 48.5 & -0.00027 \\ 
& Lateral Ventricle SD & 48 & -0.00025 & 46.5 & -0.00024 \\ 
& 4th Ventricle Median & 49 & -0.00025 & 45.5 & -0.00024 \\ 
& Inferior Lateral Ventricle Median & 50 & -0.00026 & 43.5 & -0.00021 \\ 
& Superior Temporal Median & 51 & -0.00026 & 47 & -0.00023 \\ 
& Paracentral SD & 52 & -0.00029 & 42.5 & -0.00020 \\ 
& Cuneus Median & 53 & -0.00033 & 52.5 & -0.00033 \\ 
& Thalamus Proper Median & 54 & -0.00035 & 43.5 & -0.00022 \\ 
& Inferior Temporal Median & 55 & -0.00047 & 55 & -0.00041 \\ 
& Middle Temporal Median & 56 & -0.00052 & 56 & -0.00046 \\ 
\bottomrule
\end{tabular}
\caption{Feature importance scores and rankings from full data and subsamples for all 56 high-variance VoxelDBM regions.}
\label{tab:all_regions_loco}
\end{table}

\begin{revision}
\subsection{LOCO-MP under gradient boosting machines}
\label{sec:gbm}

Though we primarily study the performance of LOCO-MP under nonparametric random survival forests, the feature selection framework may utilize other machine learning survival algorithms, such as gradient boosting machines (GBM). We incorporate a survival GBM implementation~\cite{gbm_pack} into LOCO-MP (GBM-LOCO), and capture its DBM feature importance ranks in OPERA I across ten 80\% subsamples, analogously to the experiments in Section \ref{sec:feat-imp}. We also capture the feature ranks of GBM alone, which are assigned based on the reduction in negative Cox partial log likelihood.

In Table \ref{tab:gbm_loco}, for the top six LOCO-MP random forest features, we present the ranks of their median predictive importance across ten subsamples under GBM and GBM-LOCO. The importance of the third and lateral ventricles and the cerebellum white matter is retained for both GBM and GBM-LOCO. The GBM-LOCO ranks are more concordant with RF-LOCO than are the GBM ranks, particularly for the third ventricle and precuneus. The parahippocampal region is unimportant under both GBM and GBM-LOCO. This suggests that an ensembling approach of multiple machine methods under LOCO could help further validate brain regions that are stably identified.

\begin{table}[ht]
\footnotesize
\centering
\begin{tabular}{rllcc}
\toprule
& \textbf{Region} & \textbf{Feature Summary} & \multicolumn{2}{c}{\textbf{Median Subsample Rank}} \\ 
\cmidrule(lr){4-5}
& & \multicolumn{1}{c}{\textbf{Statistic}} & \textbf{GBM} & \textbf{GBM-LOCO} \\ 
\midrule
& 3rd Ventricle & \multicolumn{1}{c}{Std. Dev} & 6 & 2  \\ 
& Lateral Ventricle & \multicolumn{1}{c}{Median} & 1 & 1 \\ 
& Precuneus & \multicolumn{1}{c}{Median} & 18.5 & 7\\ 
& Cerebellum White Matter & \multicolumn{1}{c}{Median} & 5 & 5 \\ 
& Lingual & \multicolumn{1}{c}{Median} & 9.5 & 13 \\ 
& Parahippocampal & \multicolumn{1}{c}{Median} & 25 & 21.5 \\ 
\bottomrule
\end{tabular}
\caption{Ranks of median predictive importance across ten subsamples under GBM and GBM-LOCO.  Results presented for top six LOCO-MP random survival forest features.}
\label{tab:gbm_loco}
\end{table}


\end{revision}

\section{Penalized Cox proportional hazards modeling}
\label{sec:cox-models}
\begin{figure}[h]
    \centering
    \includegraphics[width=0.85\linewidth]{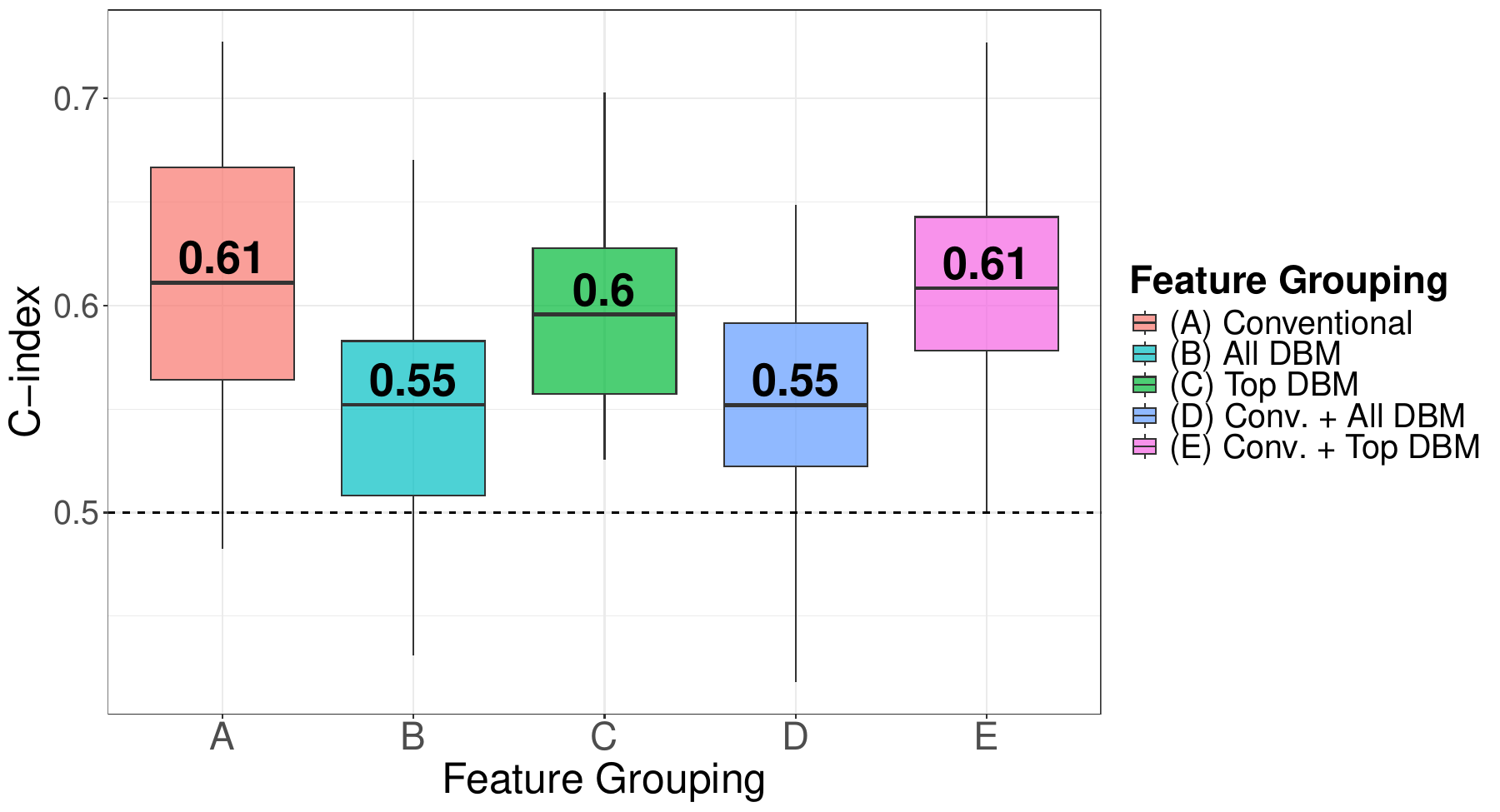}
    \caption{CCDP24 Cox model results (ridge). Boxplots show the distribution of penalized Cox proportional hazards model (ridge penalty) test-set C-indices across all cross-validation folds and repeats for outcome {CCDP24}. The $x$-axis consists of the five feature groupings and the $y$-axis shows the test set C-indices. ``Top MRI" refers to the top six DBM features. The annotated C-index represents the median value for a specific feature grouping.}
    \label{fig:survmod-cox-ccdp}
\end{figure}

\begin{figure}[h]
    \centering
    \includegraphics[width=0.85\linewidth]{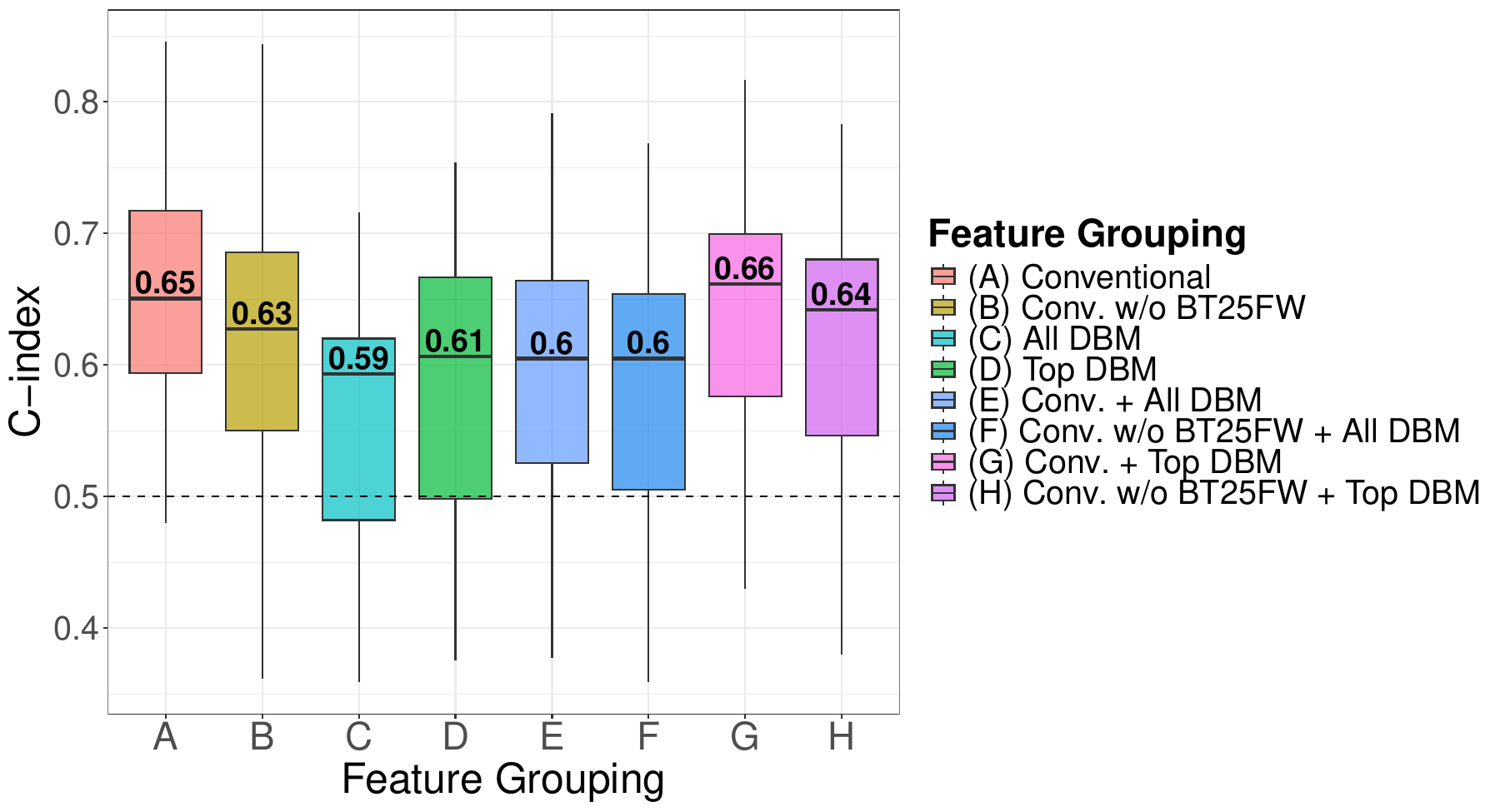}
    \caption{S25FW Cox model results (ridge). Boxplots show the distribution of penalized Cox proportional hazards model (ridge penalty) test-set C-indices across all cross-validation folds and repeats for outcome {25FW}. The $x$-axis consists of the five feature groupings and the $y$-axis shows the test set C-indices. ``Top MRI" refers to the top six DBM features. The annotated C-index represents the median value for a specific feature grouping.}
    \label{fig:survmod-cox-25fw}
\end{figure}

\begin{figure}[h]
    \centering
    \includegraphics[width=0.85\linewidth]{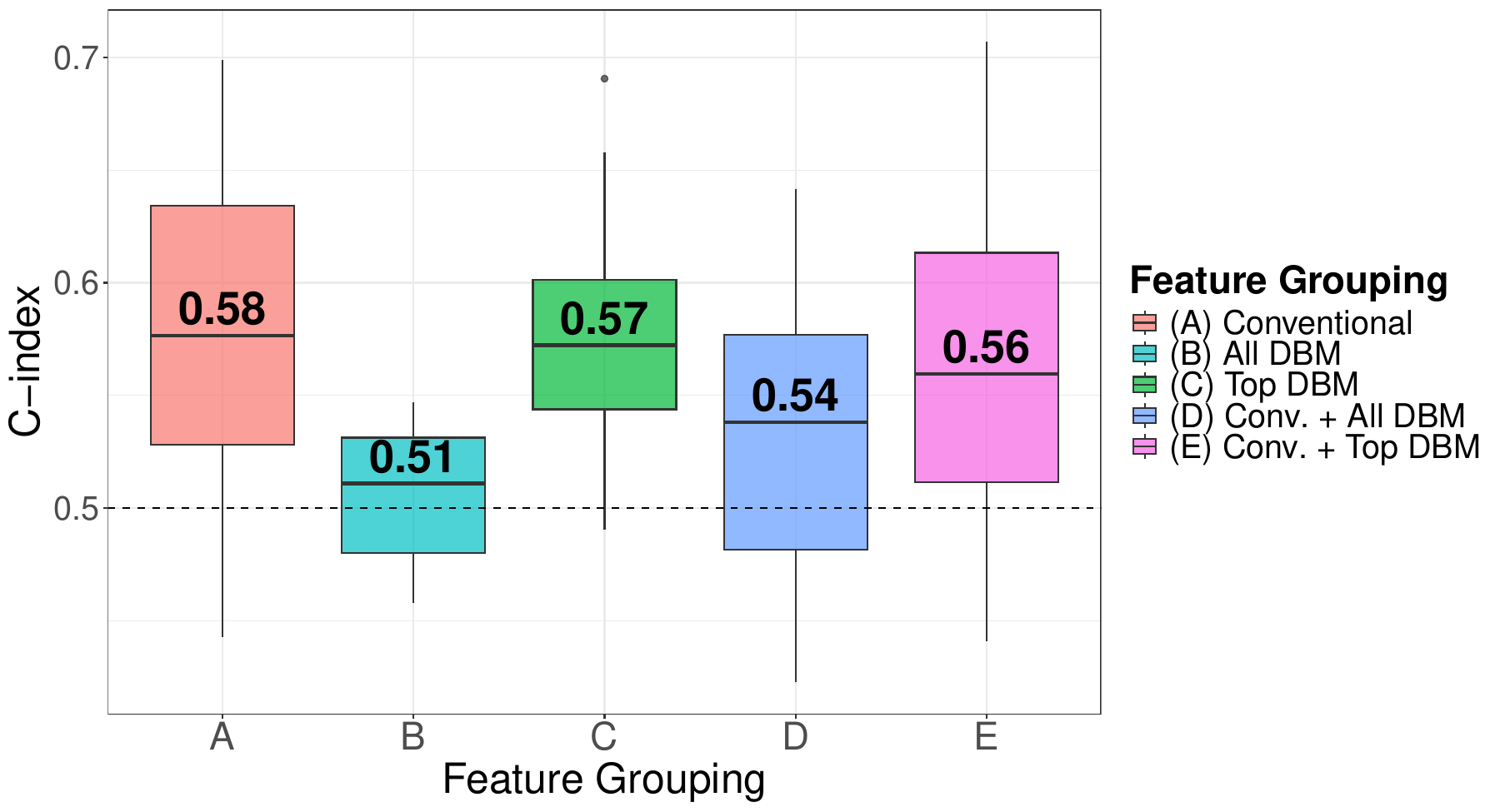}
    \caption{CCDP24 Cox model results (lasso). Boxplots show the distribution of penalized Cox proportional hazards model (lasso penalty) test-set C-indices across all cross-validation folds and repeats for outcome {CCDP24}. The $x$-axis consists of the five feature groupings and the $y$-axis shows the test set C-indices. ``Top MRI" refers to the top six DBM features. The annotated C-index represents the median value for a specific feature grouping.}
    \label{fig:survmod-lasso-ccdp}
\end{figure}

\begin{figure}[h]
    \centering
    \includegraphics[width=0.85\linewidth]{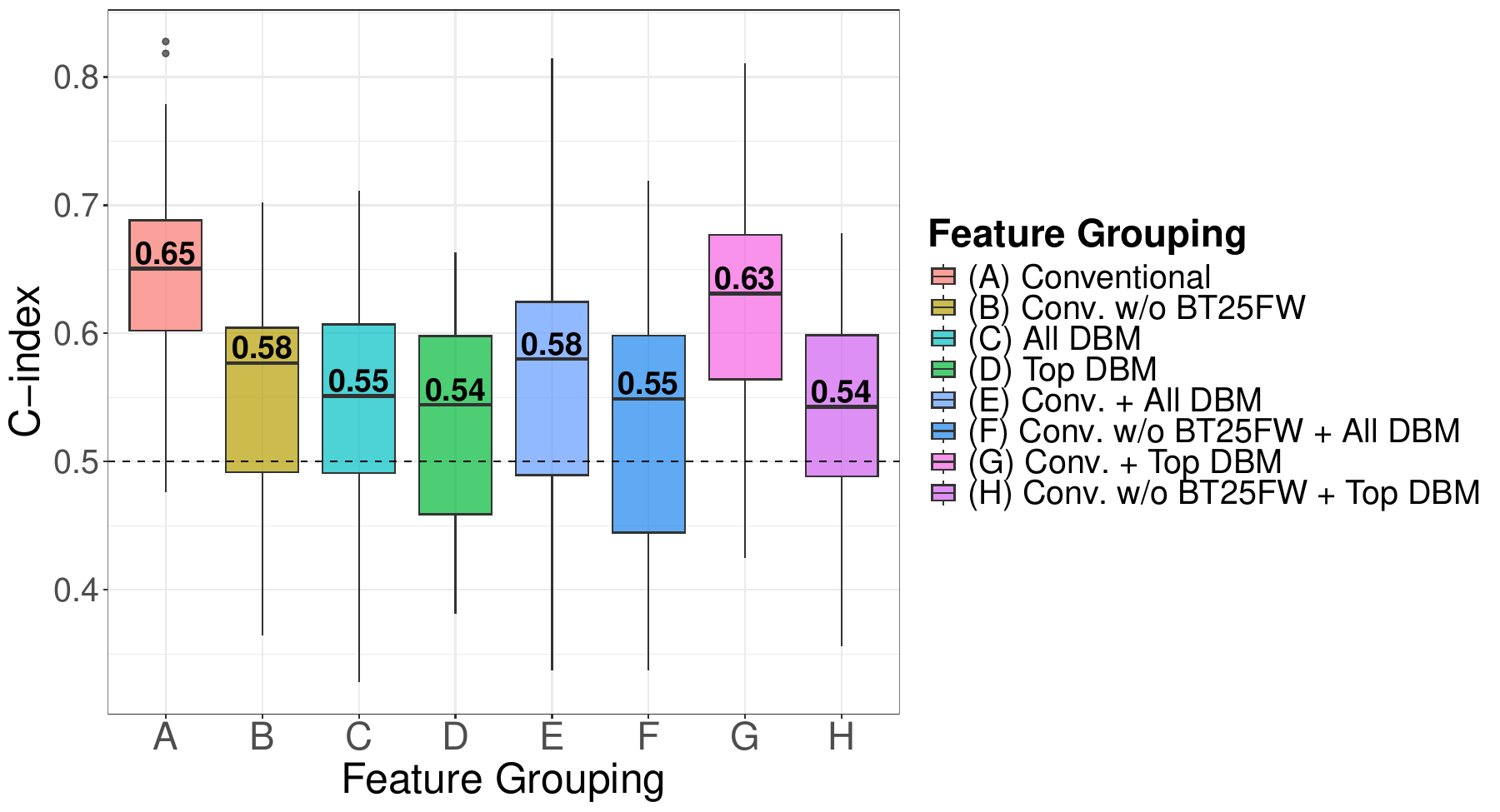}
    \caption{S25FW Cox model results (lasso). Boxplots show the distribution of penalized Cox proportional hazards model (lasso penalty) test-set C-indices across all cross-validation folds and repeats for outcome {25FW}. The $x$-axis consists of the five feature groupings and the $y$-axis shows the test set C-indices. ``Top MRI" refers to the top six DBM features. The annotated C-index represents the median value for a specific feature grouping.}
    \label{fig:survmod-lasso-25fw}
\end{figure}

Figures \ref{fig:survmod-cox-ccdp} and \ref{fig:survmod-cox-25fw} show the distribution of test set C-indices across all folds for each feature group when a penalized Cox proportional hazards model (ridge penalty) is used to predict progression. Figure \ref{fig:survmod-cox-ccdp} uses the CCDP24 outcome and Figure \ref{fig:survmod-cox-25fw} uses the 25-foot walking outcome. \smallrevision{Figures \ref{fig:survmod-lasso-ccdp} and \ref{fig:survmod-lasso-25fw} show the analogous plots for the Cox proportional hazards model with a lasso penalty.}

Figures \ref{fig:ridge-ccdp-multitop} and \ref{fig:ridge-25fw-multitop} show the stability of the test set C-indices across varying numbers of top LOCO-MP features for the penalized Cox proportional hazards model (ridge penalty). Figure \ref{fig:ridge-ccdp-multitop} uses the CCDP24 outcome and Figure \ref{fig:ridge-25fw-multitop} uses the 25FW outcome. \smallrevision{Again, Figures X and Y show the analogous plots for the Cox proportional hazards model with a lasso penalty.}

\begin{figure}[h]
    \centering
    \includegraphics[width=0.85\linewidth]{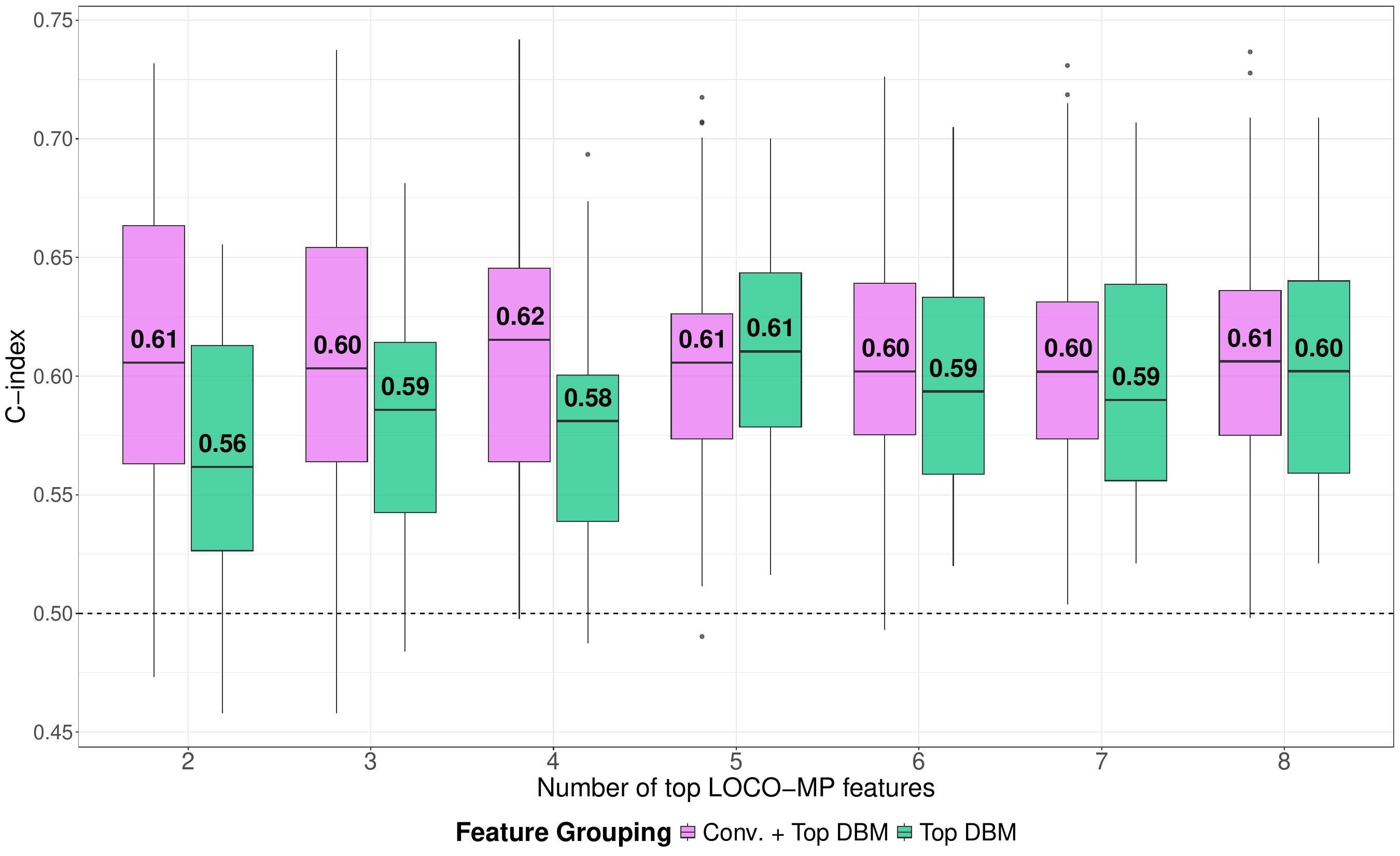}
    \caption{CCDP24 Cox model stability (ridge). Boxplots show the distribution of CCDP24 test-set C-indices for the conventional + top MRI and top MRI-only Cox proportional hazard (ridge penalty) models, plotted across varying numbers of top LOCO-MP features. The $x$-axis represents the number of top LOCO-MP features selected and the $y$-axis is the C-index.}
    \label{fig:ridge-ccdp-multitop}
\end{figure}

\begin{figure}[h]
    \centering
    \includegraphics[width=0.85\linewidth]{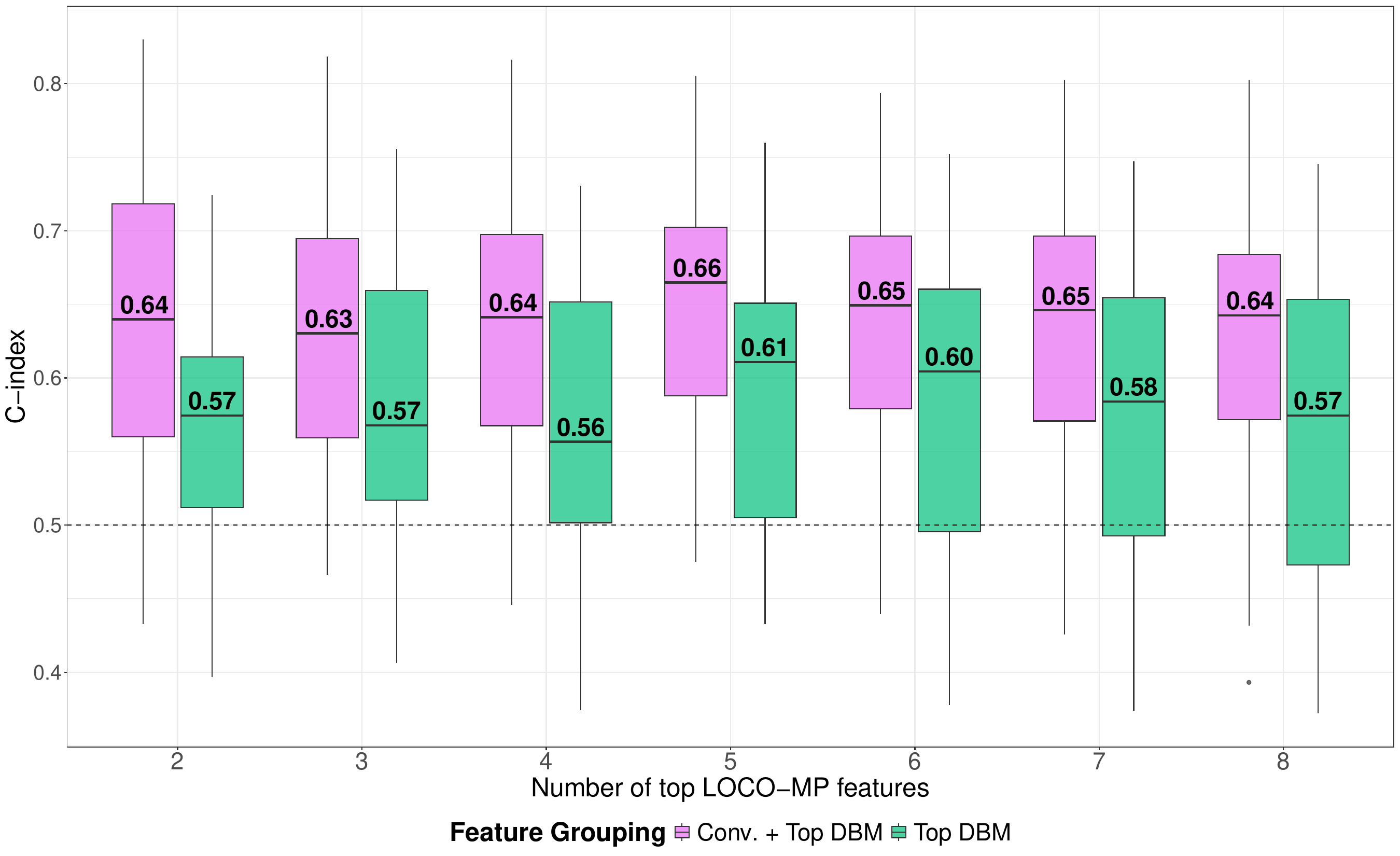}
    \caption{S25FW Cox model stability (ridge). Boxplots show the distribution of S25FW test-set C-indices for the conventional + top MRI and top MRI-only Cox proportional hazards (ridge penalty) models, plotted across varying numbers of top LOCO-MP features. The $x$-axis represents the number of top LOCO-MP features selected and the $y$-axis is the C-index.}
    \label{fig:ridge-25fw-multitop}
\end{figure}

\begin{figure}[h]
    \centering
    \includegraphics[width=0.85\linewidth]{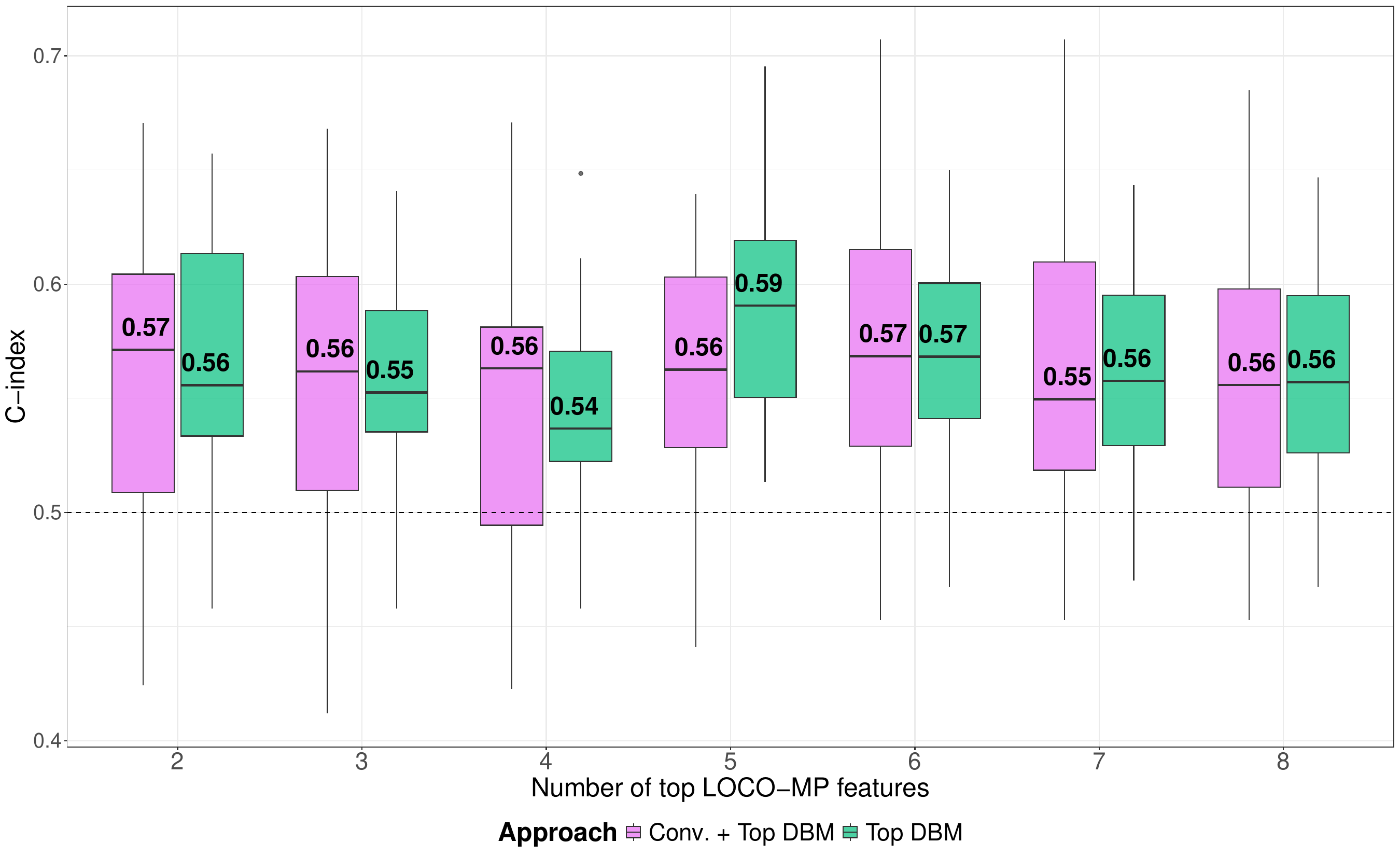}
    \caption{CCDP24 Cox model stability (lasso). Boxplots show the distribution of CCDP24 test-set C-indices for the conventional + top MRI and top MRI-only Cox proportional hazard (lasso penalty) models, plotted across varying numbers of top LOCO-MP features. The $x$-axis represents the number of top LOCO-MP features selected and the $y$-axis is the C-index.}
    \label{fig:ridge-ccdp-multitop}
\end{figure}

\begin{figure}[h]
    \centering
    \includegraphics[width=0.85\linewidth]{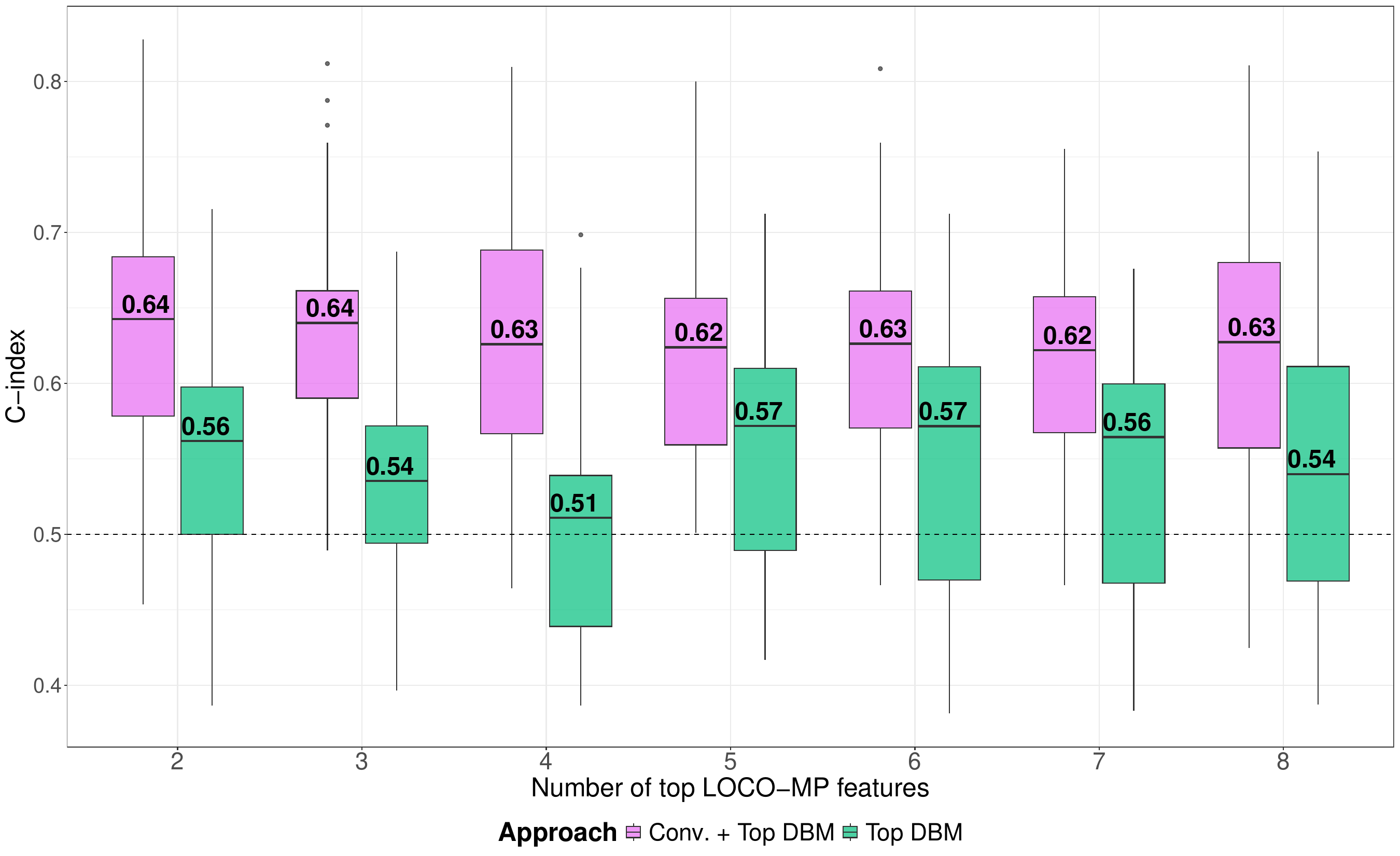}
    \caption{S25FW Cox model stability (lasso). Boxplots show the distribution of S25FW test-set C-indices for the conventional + top MRI and top MRI-only Cox proportional hazards (lasso penalty) models, plotted across varying numbers of top LOCO-MP features. The $x$-axis represents the number of top LOCO-MP features selected and the $y$-axis is the C-index.}
    \label{fig:ridge-25fw-multitop}
\end{figure}

\begin{revision}
\section{Generalizability across study arms} \label{sec:generalizability}
To evaluate the generalizability of the LOCO-MP features, we replicated the random survival forest models in Section \ref{sec:survmod-results} using an additional internal  control cohort that is demographically matched to OPERA I (OPERA II: NCT01412333)~\cite{hauser2017ocrelizumab}. This arm was previously not used for feature selection or modeling and serves as an independent validation cohort.  We use the same feature groupings and modeling process outlined in Section \ref{sec:classical-survmod} to assess whether the LOCO-MP features identified with OPERA I can generalize to an unseen cohort. The RSF models are retrained on OPERA II for CCDP24 and S25FW progression with and without the conventional features, with hyperparameter tuning for number of trees. We compare models using all DBM features versus the LOCO-MP selected top DBM features, and look at median performance over thirty 80/20 data splits, analogously to the main text. Table \ref{tab:opera2_deltas} shows that the Top DBM features induce substantial C-Index gains over all DBM features for each retrained model set.

\begin{table}[ht]
\footnotesize
\centering
\begin{tabular}{rllcc}
\toprule
&  & \textbf{Includes} & \textbf{Median C-Index $\Delta$} \\
& \textbf{Outcome} & \textbf{Conventional Features} & \textbf{of Top DBM} \\

\midrule
& CCDP & No & +0.06  \\ 
& CCDP & Yes & +0.04 \\ 
& S25FW & No & +0.15  \\ 
& S25FW & Yes & +0.12  \\ 
\bottomrule
\end{tabular}
\caption{Median C-Index improvements when using sparse DBM feature set over all DBM features in OPERA II validation cohort ($n=340$). Results are from thirty 80/20 splits of the data.}
\label{tab:opera2_deltas}
\end{table}

\end{revision}







\end{document}